\documentclass[preprintnumbers, floatfix,letterpaper,aps,prd,epsfig,nofootinbib,
twocolumn
]{revtex4-1}
\pdfoutput=1
\usepackage{bm,graphicx,dcolumn,epstopdf,epsf, latexsym,mathbbol, amssymb,amsmath,color,slashed, mathrsfs,mathcomp, simplewick}
\usepackage[marginal]{footmisc}
\usepackage[center]{subfigure}
\usepackage{multirow}
\usepackage{makecell}
\usepackage[colorlinks,linkcolor=blue,citecolor=blue,urlcolor=blue]{hyperref}

\begin{document}
	\allowdisplaybreaks
	\newcommand{\bq}{\begin{equation}}
	\newcommand{\eq}{\end{equation}}
	\newcommand{\bqn}{\begin{eqnarray}}
	\newcommand{\eqn}{\end{eqnarray}}
	\newcommand{\nb}{\nonumber}
	\newcommand{\lb}{\label}
	\newcommand{\f}{\frac}
	\newcommand{\p}{\partial}
	\newcommand{\PRL}{Phys. Rev. Lett.}
	\newcommand{\PLB}{Phys. Lett. B}
	\newcommand{\PRD}{Phys. Rev. D}
	\newcommand{\CQG}{Class. Quantum Grav.}
	\newcommand{\JCAP}{J. Cosmol. Astropart. Phys.}
	\newcommand{\JHEP}{J. High. Energy. Phys.}
	\newcommand{\red}{\textcolor{black}}
	
	\title{Shadow and Quasinormal Modes of a Rotating Loop Quantum Black Hole}
	
	\author{Cheng Liu${}^{a, b}$}
	
	\author{Tao Zhu${}^{a, b}$}
	\email{zhut05@zjut.edu.cn; Corresponding author}
	
	\author{Qiang Wu${}^{a, b}$}
	
	\author{Kimet Jusufi${}^{c, d}$}
	
	\author{Mubasher Jamil${}^{a,b, e}$ }

	\author{Mustapha Azreg-A\"{i}nou${}^{f}$}

	\author{Anzhong Wang${}^{g}$}
	
	\affiliation{${}^{a}$Institute for Theoretical Physics \& Cosmology, Zhejiang University of Technology, Hangzhou, 310023, China\\
		${}^{b}$ United Center for Gravitational Wave Physics (UCGWP),  Zhejiang University of Technology, Hangzhou, 310023, China\\
		${}^c$ Physics Department, State University of Tetovo, Ilinden Street nn, 1200, Tetovo, North Macedonia \\
		${}^d$ Institute of Physics, Faculty of Natural Sciences and Mathematics, Ss. Cyril and Methodius University, Arhimedova 3, 1000 Skopje, North Macedonia \\
		${}^{e}$ School of Natural Sciences, National University of Sciences and Technology, Islamabad, 44000, Pakistan \\
		${}^{f}$ Ba\c{s}kent University, Engineering Faculty, Ba\u{g}l{\i}ca Campus, 06790-Ankara, Turkey\\
		${}^{g}$ GCAP-CASPER, Physics Department, Baylor University, Waco, Texas 76798-7316, USA}

	\date{\today}
	
\begin{abstract}
		
In this paper,  we construct an effective rotating loop quantum black hole (LQBH) solution, starting from the spherical symmetric LQBH by applying the Newman-Janis algorithm modified by Azreg-A\"{i}nou's non-complexification procedure, and study the effects of  loop quantum gravity  { (LQG) on its shadow}. Given the rotating  {LQBH}, we discuss its horizon, ergosurface, and regularity  {as} $r \to 0$. Depending on the values of the specific angular momentum $a$ and the polymeric function $P$ arising from  {LQG}, we  {find} that the rotating solution we obtained can represent a regular black hole, a regular extreme black hole, or  a regular  spacetime  {without horizon (a non-black-hole solution)}.  We also {study} the effects of {LQG} and rotation, and {show} that,  in addition to the specific angular momentum, the polymeric function {also} causes deformations in the size and shape of the black hole shadow. Interestingly, for a given value of $a$ and inclination angle $\theta_0$, the apparent size of the shadow monotonically decreases,  and the shadow gets more distorted with increasing $P$. We also {consider the effects of  $P$ on the deviations from the circularity of the shadow, and find} that the deviation from circularity increases with increasing $P$ for fixed values of $a$ and $\theta_0$. Additionally,  we explore the observational implications of $P$ in comparing with the latest Event Horizon Telescope (EHT) observation of the supermassive black hole, M$87$. The connection between the shadow radius and quasinormal modes in the eikonal limit as well as {the} deflection of massive particles are also considered.
		
\end{abstract}

\maketitle
	
\section{Introduction}
	
	Recently, the EHT Collaboration announced their first image concerning the detection of an event horizon of a supermassive black hole at the center of a neighboring elliptical M87 galaxy \cite{m87, Akiyama:2019brx, Akiyama:2019eap, Akiyama:2019bqs, Akiyama:2019fyp, Akiyama:2019sww}. With this image, it {was found} that the diameter of the center black hole shadow is $(42\pm 3) \; {\rm \mu as}$ with a deviation $\lesssim 10 \%$ from circularity, which leads to a measurement of the center {mass,} $M=(6.5\pm 0.7)\times 10^9 M_{\odot}$ \cite{m87}. These results are in good agreement with the predictions of general relativity by assuming the geometry of black hole is described by the Kerr metric. Future improvement{s} of the observations, such as those {from} the Next Generation Very Large Array \cite{ngVLA}, the Thirty Meter Telescope \cite{TMT}, and BlackHoleCam \cite{BlackHoleCam}, can provide unique opportunities for us to get deep insight on the nature of  {black hole spacetimes} in the regime of strong gravity. More importantly, these precise observations could provide a { valuable}  window  to explore, distinguish, or constrain physically viable black hole solutions that exhibit small {deviations from the} Kerr metric.
	
	A black hole shadow is a two-dimensional dark zone in the celestial sphere caused by the strong gravity of the black hole. The shape and size of the shadow mainly depend on the geometry of the black hole spacetime. With this reason, by observing both the shape and the size of the shadow, one is able to extract valuable information about the black hole spacetime.   The theory of black hole shadows {has been  under investigations for decades, and now is well developed}. The calculations of the shadow size and shape,   {and their observational implications from different black hole spacetimes or spacetimes of compact objects  with exotic matters ether in  general relativity or in modified theories of gravity} have been extensively {studied, see, for example,}  \cite{Claudel:2000yi, wang_shadow_2019, gb, kz, Younsi:2016azx, cs, wei_observing_2013, Amir:2018pcu, pfdm, kds, mog, wang_shadows_2019, rr, Dastan:2016bfy, ks, nc, Bambi:2010hf, Jusufi:2019nrn, hou_black_2018, konoplya_shadow_2019, Haroon:2019new, ns, Bambi:2008jg, wei_intrinsic_2019, wei_curvature_2019, Zhu:2019ura, Li:2020zxi, Li:2019lsm, Konoplya:2019xmn, Allahyari:2019jqz, Dokuchaev:2019jqq, Jusufi:2019caq, Ding:2019mal, Tian:2019yhn, Konoplya:2019fpy, Konoplya:2019goy, Vagnozzi:2020quf, Yu:2020bxd, Kumar:2020hgm, Azreg:2015}. Recently,  the first observational data of the shadow image captured by EHT \cite{m87, Akiyama:2019brx, Akiyama:2019eap, Akiyama:2019bqs, Akiyama:2019fyp, Akiyama:2019sww} have been used for constraining the black hole parameters and deviations from the Kerr metric \cite{Bambi:2019tjh, Vagnozzi:2019apd, Banerjee:2019nnj, Zhu:2019ura, Held:2019xde}. Although these constraints are still not stringent enough, these works do show that the observational data   does have the capacity for constraining black hole parameters beyond those {presented} in GR. One expects that future precise observations can make more significant constraints.
	
	On the other hand, the strong gravity regime near a black hole is also thought to be a region that may help us to explore the quantum nature of the spacetime.
	Recently, in the context of {loop quantum gravity (LQG)}, a spherical symmetric black hole,  known as LQBH or self-dual black hole, was constructed \cite{LQG_BH}
	\footnote{{In the last couple of years, LQBHs have been extensively studied, see, for instance, \cite{AOS18a,AOS18b,BBy18,ABP19,ADL20}.  For more details, we refer the reader to
	the review articles, \cite{Perez17,Rovelli18,BMM18,Ashtekar20}.}}. Several phenomenological implication{s of this black hole have been investigated} \cite{Alesci:2011wn, Chen:2011zzi, Dasgupta:2012nk, Hossenfelder:2012tc, Barrau:2014yka, Sahu:2015dea, Cruz:2015bcj}. {In particular, a} rotating LQBH has been constructed by using the Newman-Jains procedure based on the spherical symmetric LQBH \cite{Caravelli:2010ff}. However, it was pointed out  \cite{AzregAinou:2011fq} that the { construction of the rotating LQBH presented}  in \cite{Caravelli:2010ff} is not valid. Thus it is still an open question for generating rotating LQBH {solutions from} the  spherical symmetrical LQBH by using the Newman-Janis algorithm \cite{NJ1, NJ2, Til, drake}.  In this algorithm, one of {the} critical steps   {is how to complexify the radial coordinate $r$. In fact, there does not exist a standard role to follow,  and the  complexification is rather arbitrary} \cite{A}. {Recently, one of us has proposed  a new procedure~\cite{A, A1}, in which one simply} drops the complexification step of the Newman-Janis algorithm. The procedure has {been applied to  various cases} ~\cite{Jusufi:2019nrn, A, A1, s1, s2, s3, s4, s5, s6, s7, s8, s9, s10, s11, s12, s13, s14, s15, s16, s17, Jusufi:2019caq, s18}. One of {the} main purposes of this paper is to construct the rotating LQBH solution by using this new procedure, and then to investigate its main geometric properties and the effects of LQG on the shadow of the rotating LQBH solution.
	
In addition, it has been explored recently that there may exist interesting connections among different physical phenomenas arising from the strong gravity regime in black holes. In \cite{cardoso}, it was argued by Cardoso {\em et. al.} that the black hole quasinormal modes (QNMs) in the eikonal limit can be connected to the last circular null geodesic, in which the real part of the QNMs is related to the angular velocity of the last circular null geodesic, while the imaginary part is related to the Lyapunov exponent that determines the instability time scale of the orbit. Later on, Stefanov {\em et. al.} \cite{Stefanov:2010xz} found another connection between black hole QNMs in the eikonal limit and gravitational lensing in the strong deflection limit. Since both the last circular null geodesic and the gravitational lensing in the strong deflection limit are closely related to the boundary of shadow of a black hole, it is natural to ask whether there is a connection between the black hole QNMs and black hole shadow. Recently, such connection has been explored by Jusufi \cite{Jusufi:2019ltj}, in which a relation between the real part of the black hole QNMs and shadow radius has been established \footnote{It is has been also shown in \cite{wei2019} that the angular velocity of the last circular null geodesic can be related to the photon sphere of the black hole.}. In this paper, we would like to extend and apply these works to the LQBH by studying the effects of LQG on the QNMs of the LQBH solution.

	
	The plan of our paper is as follows. In Sec. II, starting from the spherical symmetric LQBH, we generate  {a}  rotating LQBH by using the Newman-Janis procedure without the complexification {of the radial coordinate}.
	 {In Sec. III, we study the physical properties of its horizon, ergosurface, and regularity in the limit} $r \to 0$. Then, in Sec. IV,  with this regular rotating black hole solution,  we study the null geodesic equations and orbital equations of photons. In Sec. V, {we study the shadow of the rotating LQBH and discuss its size and shape allowing us to investigate  the effects of LQG}.  The connection between the shadow radius and quasinormal modes is presented in Sec. VI. We also derive the deflection of massive particles in this LQBH spacetime in Sec. VII. {Our main conclusion are} presented in Sec. VIII.

\section{A rotating loop black hole \label{secrot}}
	
	We start with the  effective {LQG-corrected} geodesically complete Schwarzschild metric, which can be expressed in the form \cite{LQG_BH}
	\begin{equation}\label{1}
	ds^2=f(r)dt^2-\frac{dr^2}{g(r)}-h(r)(d\theta^2+\sin^2\theta d\phi^2),
	\end{equation}
	where the metric functions $f(r)$, $g(r)$, and $h(r)$ are given by
	\begin{eqnarray}
	f(r)&=&\frac{(r-r_+)(r-r_-)(r+r_*)^2}{r^4+a_0^2},\nonumber\\
	g(r)&=&\frac{(r-r_+)(r-r_-)r^4}{(r+r_*)^2(r^4+a_0^2)},\nonumber\\
	h(r)&=&r^2+\frac{a_0^2}{r^2}.
	\end{eqnarray}
	Here $r_+=2 M/(1+P)^2$ and $r_{-} = 2 M P^2/(1+P)^2$ are the two horizons, and $r_{*}= \sqrt{r_+ r_-} = 2 MP/(1+P)^2$ with $M$ denoting the ADM mass of the solution and $P$ being the polymeric function $P=(\sqrt{1+\epsilon^2}-1)/(\sqrt{1+\epsilon^2}+1)$,  where $\epsilon$ denotes a product of the Immirzi parameter $\gamma$ and the polymeric parameter $\delta$, i.e., $\epsilon=\gamma \delta \ll 1$. The parameter $a_0 = A_{\rm min}/8\pi$ is the minimum area gap of LQG.
	By taking $a_0=0=P$, it is easy to {see} that the above solution  reduces to the Schwarzschild black hole exactly.
	
	{Given the above LQG-corrected Schwarzschild metric, we aim to construct its rotating LQBH counterpart by using the Newman-Janis algorithm modified by Azreg-A\"{i}nou's non-complexification procedure~\cite{A, A1}}. The first step of the Newman-Janis algorithm is to transform from  {the} Boyer-Lindqiust coordinates $(t, r, \theta, \phi)$ to {the} Eddington-Finkelstein coordinates $(u, r, \theta, \phi)$. On applying the coordinate transformation $dt=du+dr/\sqrt{f(r)g(r)}$ to Eq.~(\ref{1}), we obtain
	\begin{eqnarray}\label{b}
	ds^{2}&=&f(r)du^{2}+2\sqrt{\frac{f(r)}{g(r)}}dudr-h(r)d\theta^{2}\nb\\
	&& -h(r)\sin^{2}\theta{d\phi}^2.
	\end{eqnarray}
	{In terms of null tetrads~\cite{A}, this metric can be represented as},
	\begin{eqnarray}\label{d}
	g^{ab}=l^{a}n^{b}+l^{b}n^{a}-m^{a}\bar{m}^{b}-m^{b}\bar{m}^{a},
	\end{eqnarray}
	where
	\begin{eqnarray}
	l^{a}&=&\delta^{a}_{r},\nonumber\\
	\label{e0} n^a&=&\sqrt\frac{g(r)}{f(r)}\delta^{a}_{u}-\frac{g(r)}{2}\delta^{a}_{r},\\
	m^a&=&\frac{1}{\sqrt{2h(r)}}(\delta^{a}_{\theta}+\frac{\dot{\iota}}{\sin\theta}\delta^{a}_{\phi}).\nonumber
	\end{eqnarray}
	These null tetrads satisfy the following conditions,
	\begin{eqnarray}\nonumber
	l^{a}l_{a}=n^{a}n_{a}=m^{a}m_{a}=\bar{m}^{a}\bar{m}_{a}=0,\\
	l^{a}m_{a}=l^{a}\bar{m}_{a}=n^{a}m_{a}=n^{a}\bar{m}_{a}=0,\\\label{f}\nonumber
	l^{a}n_{a}=-m^{a}\bar{m}_{a}=1.
	\end{eqnarray}
	Now we apply the second step of the Newman-Janis algorithm which consists {of performing the complex coordinate transformations in the $(u, r)$-plane,}
	\begin{eqnarray}\nonumber
	u'\rightarrow{u-\dot{\iota}{a}\cos\theta},\\\label{g}
	r'\rightarrow{r+\dot{\iota}{a}\cos\theta},
	\end{eqnarray}
	where $a$ is the rotational parameter.
	
	The third step of the Newman-Janis algorithm consists {of} complexifying the radial coordinate $r$.  {However, as mentioned above, in principle there are infinite ways to complexify $r$} \cite{A}. One of us  {proposed a new procedure~\cite{A,A1}, in which one simply drops the complexification of $r$}. Instead,  we admit that $\delta^{\mu}_{\nu}$, in Eq.~\eqref{e0}, transform as a vector under the transformation~\eqref{g},
	 and that the functions $f(r)$, $g(r)$ and $h(r)$ transform to $F=F(r,a,\theta)$, $G=G(r,a,\theta)$ and $H=H(r,a,\theta)$,  respectively. Thus,  our new null tetrads are
	\begin{eqnarray}
	l^{a}&=&\delta^{a}_{r},\nonumber \\
	\label{e} n^a&=&\sqrt\frac{G}{F}\delta^{a}_{u}-\frac{G}{2}\delta^{a}_{r},\\\
	m^{a}&=&\frac{1}{\sqrt{2H}} \left[(\delta^{a}_{u}-\delta^{a}_{r})\dot{\iota}{a}\sin\theta+\delta^{a}_{\theta}+\frac{\dot{\iota}}{\sin\theta}\delta^{a}_{\phi}\right].\nonumber
	\end{eqnarray}
	Using these null tetrads, the contravariant components of the rotating metric are given by Eq.~(9) of Ref.~\cite{A}, which we rewrite them as \footnote{
	{In writing Eq.(\ref{MCs}),
	we used different notations from those adopted in \cite{A}. The  correspondence between the two sets of notations
	are $G(r)\to f(r)$, $F(r)\to g(r)$, and $H(r)\to h(r)$ for the nonrotating solution,  and $A(r,\theta)\to F(r,\theta)$, $B(r,\theta)\to G(r,\theta)$, and $\Psi(r,\theta)\to H(r,\theta)$ for the rotating solution.}},
	\bqn
	\lb{MCs}
	g^{uu} &=&\frac{-a^{2}\sin^{2}\theta}{H},~~~\nb\\
	g^{u\phi} &=&\frac{-a}{H}, ~~ \nb\\
	g^{ur} &=&\sqrt{\frac{G}{F}}+\frac{a^{2}\sin^{2}\theta}{H}, ~~ \nb\\
	g^{rr} &=& -G-\frac{a^{2}\sin^{2}\theta}{H}, \nb\\
	g^{r\phi} &=&\frac{a}{H},  ~~ g^{\theta\theta}=\frac{-1}{H}, \nb\\
	g^{\phi\phi} &=&-\frac{1}{H\sin^2\theta}.
	\eqn
	So the new metric in {the} Eddington-Finkelstein coordinates is,
	\begin{eqnarray}
	ds^2&=&F d u^2+2 \frac{\sqrt{F}}{\sqrt{G}} d u d r+2 a\sin ^2\theta  \left(\frac{\sqrt{F}}{\sqrt{G}}-F\right) d u d \phi \nonumber\\
	&& - 2 a\sin ^2\theta  \frac{\sqrt{F}}{\sqrt{G}} d r d \phi-H  d \theta^2  \nb\\
	&& - \sin ^2\theta  \left[H +a^2\sin ^2\theta  \left(2 \frac{\sqrt{F}}{\sqrt{G}}-F\right)\right] d \phi^2.
	\end{eqnarray}
	The final {and} crucial step is to bring this  {set of coordinates  to the Boyer-Lindquist one,} by a global coordinate transformation of the form
	\begin{eqnarray}
	du&=&dt+\lambda(r)dr,\\\nonumber
	d\phi&=&d\phi'+\chi(r)dr,
	\end{eqnarray}
	where~\cite{A,A1}
	\bqn
	\label{k}
	\lambda=-\frac{a^2+k(r)}{a^2+g(r)h(r)}, ~~ \\
	{\chi}=-\frac{a}{g(r)h(r)+a^2}, ~~ \\
	k(r)={\sqrt\frac{g(r)}{f(r)}}h(r).
	\eqn
	Since the function{s} $F$, $G$ and $H$ are still unknown, one can fix some of them to get rid of the cross term $dtdr$ in the  {above} metric. This is generally not possible in the usual Newman-Janis algorithm since these functions are fixed once the complexification of $r$ is performed, and there remains no free parameters or functions  {to achieve the transformation to the} Boyer-Lindquist coordinates.
	Now, if we choose~\cite{A,A1}
	\bqn
	\label{FG}
	F &=&\frac{\left(g(r)h(r)+a^2\cos^2\theta\right)H}{\left(k(r)+a^2\cos^2\theta\right)^2},\nb\\
	G &=&\frac{\left(g(r)h(r)+a^2\cos^2\theta\right)}{H},
	\eqn
	the rotating black hole solution in the Boyer-Lindquist coordinates turns out to be in the form of the Kerr-like metric,
	\bqn\lb{B}
	d s^2 &=&\frac{H}{k+a^2 \cos^2\theta}\Bigg[\Big(1-\frac{\sigma}{k+a^2 \cos^2\theta}\Big)d t^2 \nb\\
	&&~~~~~~~~  -\frac{k+a^2 \cos^2\theta}{\Delta}\,d r^2  +\frac{2a\sigma \sin ^2\theta}{k+a^2 \cos^2\theta}\,d td \phi \nb\\
	&&~~~~~~~~  -(k+a^2 \cos^2\theta)d \theta^2 \nb\\
	&&~~~~~~~ -\frac{[(k+a^2)^2-a^2\Delta\sin^2\theta]\sin ^2\theta}{k+a^2 \cos^2\theta}\,d \phi^2\Bigg],  \lb{mmn}
	\eqn
	where
	\bqn
	\label{D}
	\sigma(r)\equiv k-gh,~~  \Delta(r)\equiv gh+a^2,~~  k\equiv \sqrt{\frac{g(r)}{f(r)}}h(r),\nb\\
	\eqn
	and to simplify the notation we have dropped the prime from $\phi$.
	
	For the rotating LQBH, these parameters are defined as
	\begin{eqnarray}
	\Delta(r)&=& \frac{(r-r_+)(r-r_-)r^2}{(r+r_*)^2}+a^2,\\
	\sigma(r)&=& \frac{r^4+a_0^2-(r-r_+)(r-r_-)r^2}{(r+r_*)^2},\\
	k&=& \frac{r^4+a_0^2}{(r+r_*)^2},
	\end{eqnarray}
	and the metric~\eqref{B} takes the form
	\bqn
	d s^2 &=& \frac{H}{\rho^2} \left[\frac{\Delta}{\rho^2}(dt-a\sin^2\theta d\phi)^2-\frac{\rho^2}{\Delta}dr^2-\rho^2d\theta^2\right. \nb\\
	&& ~~~~ \left. -\frac{\sin^2\theta}{\rho^2}(a dt-(k+a^2)d\phi)^2 \right],  \lb{mmm}
	\eqn
	with
	\begin{equation}
	\rho^2=k+a^2\cos^2\theta,
	\end{equation}
	where $a=L/M$ is the specific angular momentum (rotation parameter),  and $M$ and $L$ are the mass and angular momentum of the black hole, respectively. It is easy to verify that when $a=0$ we recover the non-rotating LQBH solution.
	
\section{Properties of The Rotating Loop Black Hole}
	
	In this section, we discuss the physical properties of the rotating LQBH obtained in the previous section. Specifically, we will discuss the properties of the horizon, the ergosurfaces, and the regularity of the rotating LQBH.
	
	\subsection{Horizon}
	
	The event horizon  is defined by the surface $g^{rr}=0$.
	For the rotating LQBH, the existence of the horizon and their radii depend on the mass $M$, angular momentum $a$,  and the polymeric parameters $P$ arising from LQG. From the metric (\ref{mmm}) we have
	\bqn
	g_{rr}=-\frac{H}{\Delta}.
	\eqn
	Assuming $H$ to be a regular function as { its static counterpart $h=\lim_{a\to 0}H$, the component $g_{rr}$ becomes} singular when $\Delta=0$, which yields
	\bqn \lb{CON}
	\frac{(r-r_+)(r-r_-)r^2}{(r+r_*)^2}+a^2=0.
	\eqn
	The roots of the above equation gives the radii of the horizons. The nature of the roots are very sensitive to the values of $M$, $a$, and $P$. It is interesting to note that the properties of the horizon{s} is independent of the minimal area $a_0 = A_{\rm min}/8\pi$ in LQG.
	
\begin{figure}
		\centering
		\includegraphics[width=8cm]{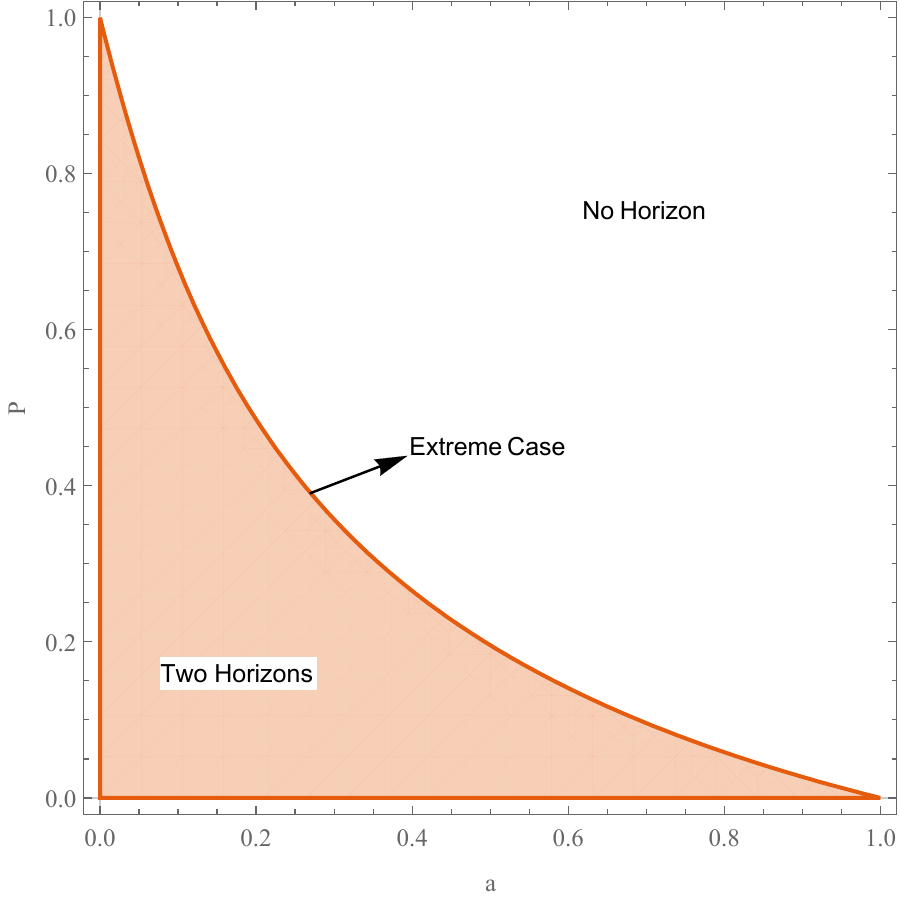}\\
		\caption{The parameter space of $P$ and $a$ for two event {horizons (orange region), one degenerate horizon (boundary between orange and white regions), and no horizon (white region), respectively}.} \label{solution}
	\end{figure}

	When Eq.~(\ref{CON}) has two {different real} roots, they correspond to the inner and outer horizons, respectively,  and represent the non-extremal rotating LQBH. {There exist particular values of the parameters $a$ and $P$ where Eq. (\ref{CON}) has a double {real} root, corresponding to an extremal black hole}. Another interesting case is that for some values of $a$ and $P$, Eq.~(\ref{CON}) does not admit any real root. {For this case, the rotating spacetime solution is termed {\textquotedblleft regular non-black-hole solution\textquotedblright}~\cite{A},  since it represents a regular spacetime solution without horizon as we shall show later. In Fig.~\ref{solution} we plot the condition for Eq.~(\ref{CON}) to have two real roots (the pink region), one double root (the curve labelled by {\textquoteleft Extreme Case\textquoteright}), and no real root (white region)}.
	
	When the black hole has two horizons, {the radii of the inner and outer horizons are obtained upon solving (\ref{CON}):}
	\bqn
	r_{h \pm}=&&\frac{1}{4} \left(r_-+r_+\right)\pm r_{h1}+r_{h2},  \lb{hor}
	\eqn
	where $r_{h1}$, $r_{h2}$ are given,  {respectively},  by (\ref{A1}) and (\ref{A2}) in Appendix A.  One can easily find that the two horizons are independent of the coordinate $\theta$.
	It is easy to check that when $P=0$, the result (\ref{hor}) reduces to the {well-known  expression} of the Kerr black hole
	\bqn
	r_{h \pm}=M \pm \sqrt{M^2-a^2}.
	\eqn
	
	\subsection{Ergosurfaces}
	
	The inner and outer ergo-surfaces are the two-dimensional surfaces which satisfy $g_{tt}=0$, {which} yields
	\bqn
	\frac{(r-r_+)(r-r_-)r^2}{(r+r_*)^2}+a^2\cos^2\theta=0.
	\eqn
	The solutions of this equation gives the radii of the ergo-surfaces for the rotating LQBH
	\bqn
	r_{e\pm}=\frac{1}{4}\left(r_++r_-\right)+r_{e1}\pm r_{e2},
	\eqn
	where $r_{e1}$ and $r_{e2}$ are given by (\ref{A6}) and (\ref{A7}) in Appendix A. Notice that for $\theta=0$ or $\pi$ the ergo-surfaces coincide with the event horizons. This property is  {valid not only for (\ref{mmm}), but also for the rotating metric (\ref{mmn}) for any given } $f$, $g$, $h$, and $H$. {The radii of ergo-surfaces of the rotating LQBH are independent of the minimal area $a_0 = A_{\rm min}/8\pi$ in LQG,  as do the radii of the inner and outer horizons.}
	
	Figs.~\ref{aa1} and \ref{bb1} {show} the behavior for the outer horizon and outer ergosurface for different values of the parameters. The orange circle {corresponds} to the static case ($a=0$) in these figures. From {them one can  find} that the shape of the outer ergo-sphere changes with the spin value $a$, and the size of the outer horizon decreases with $a$. {Similarly, for fixed $a$, the radii of the outer ergo-sphere and outer horizon decrease with increasing $P$. As $P \rightarrow 0$, the horizons and ergo-surfaces {approach  to those of the} Kerr black hole.}

	\begin{figure*}
		\centering
		\includegraphics[width=8cm]{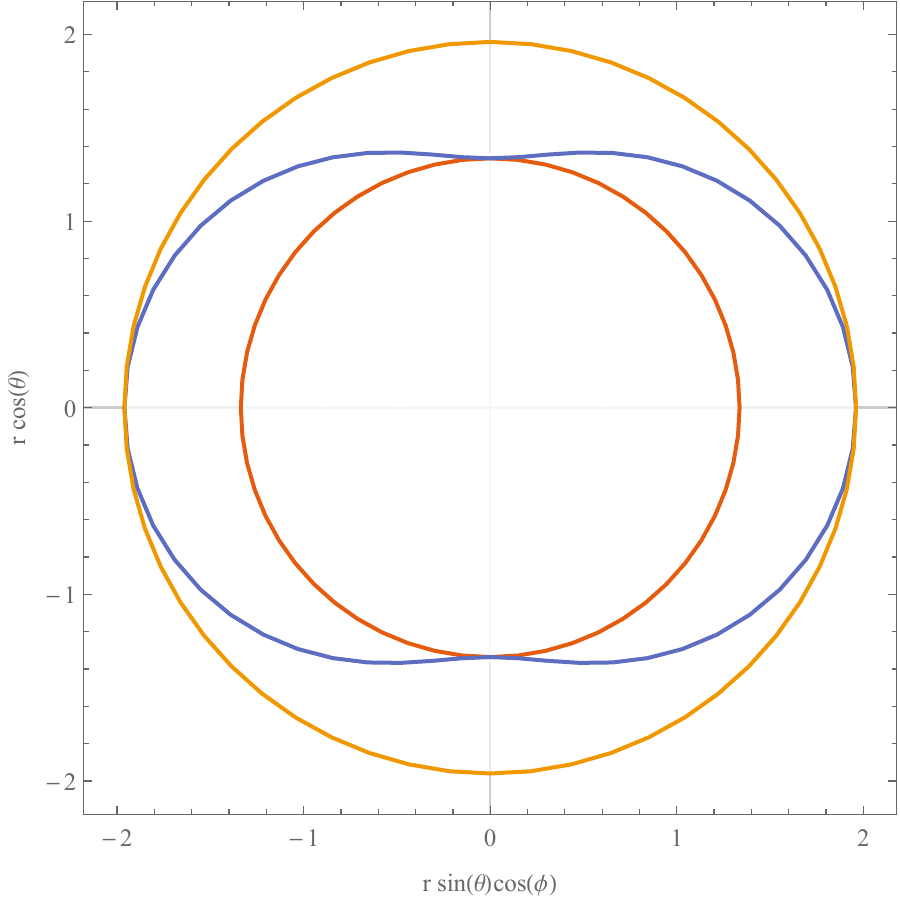}
		\includegraphics[width=8cm]{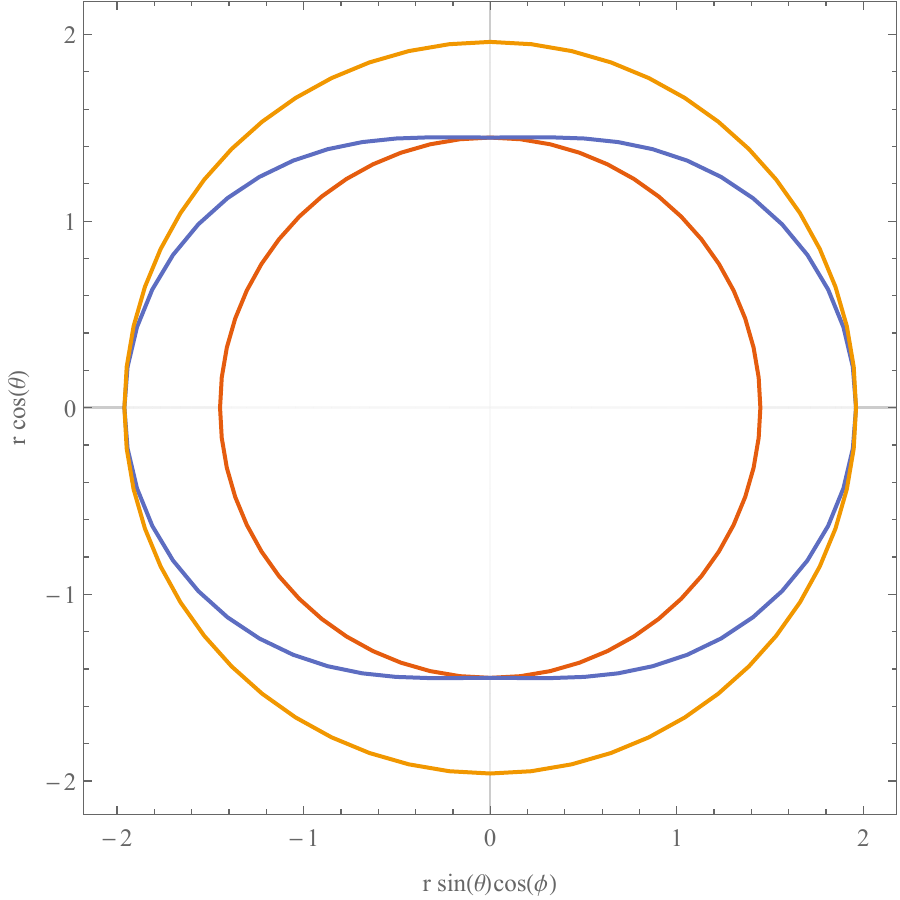}
		\includegraphics[width=8cm]{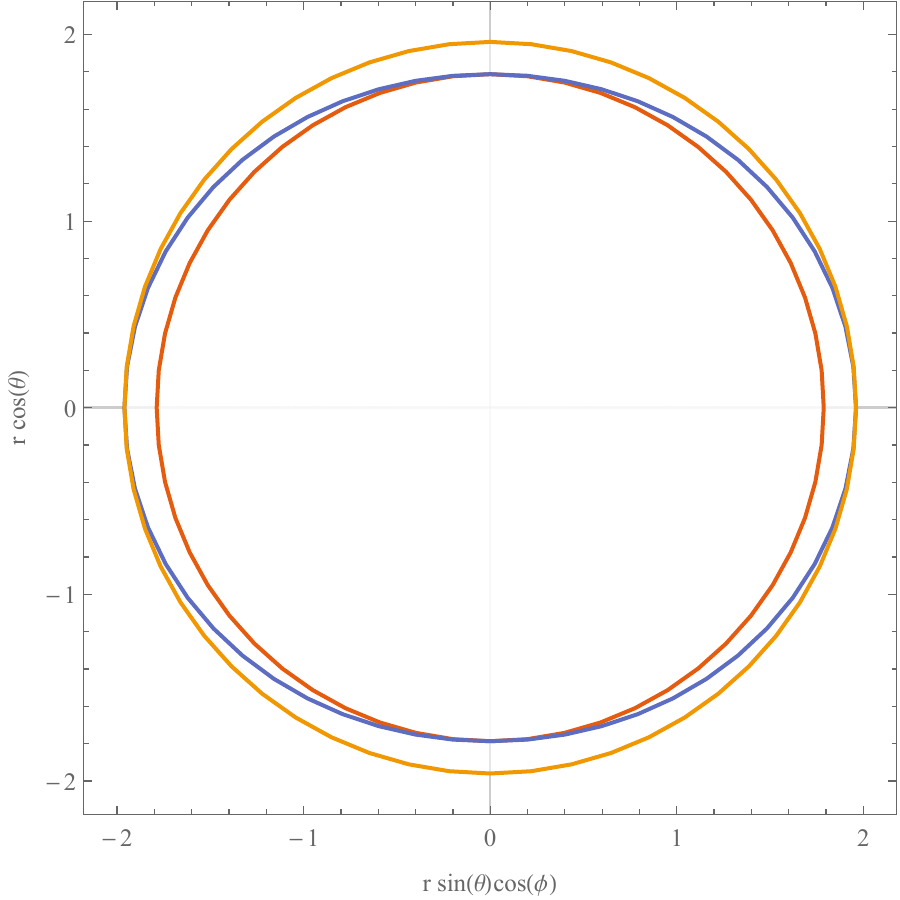}
		\includegraphics[width=8cm]{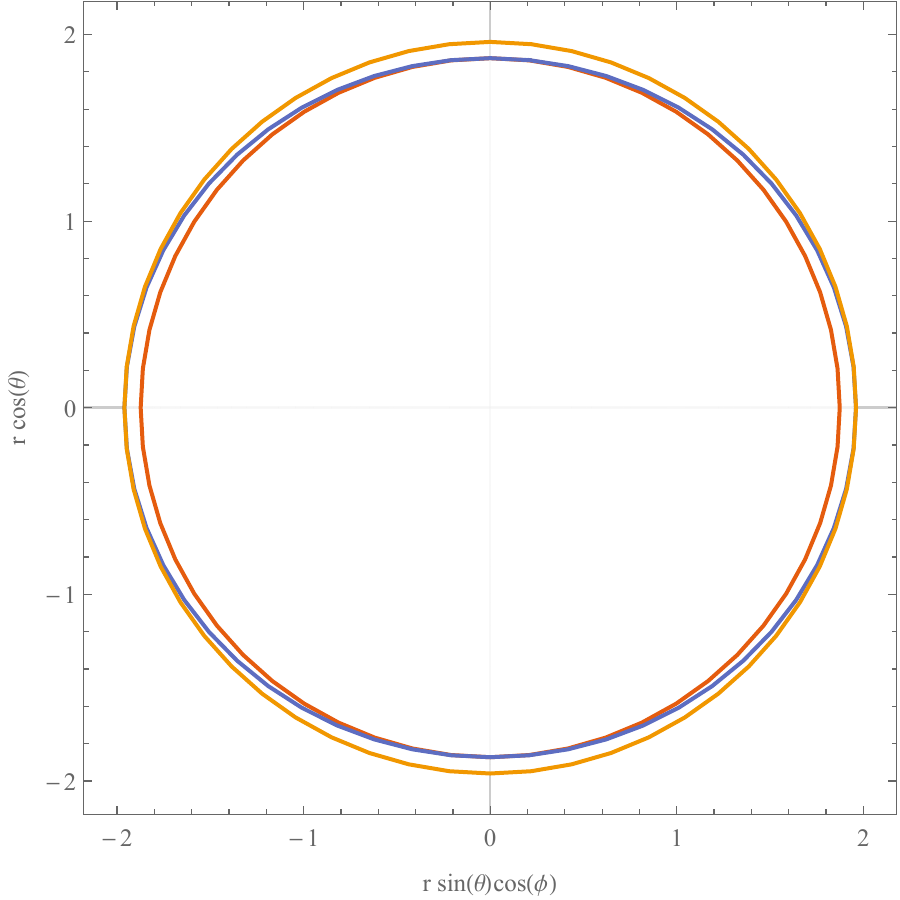}
		\caption{ Shapes of the outer horizon (red solid line) and outer ergo-sphere (blue solid line) in comparison with the outer event horizon in the static case (orange solid line) for {$M=1$} and $P=0.01$ and
		{different values of} $a$: upper and left panel: $a=0.90$; upper and right panel: $a=0.85$; bottom and left panel: $a=0.55$; and bottom and right panel: $a=0.40$.}
		\lb{aa1}
	\end{figure*}
	
	\begin{figure*}
		\centering
		\includegraphics[width=8cm]{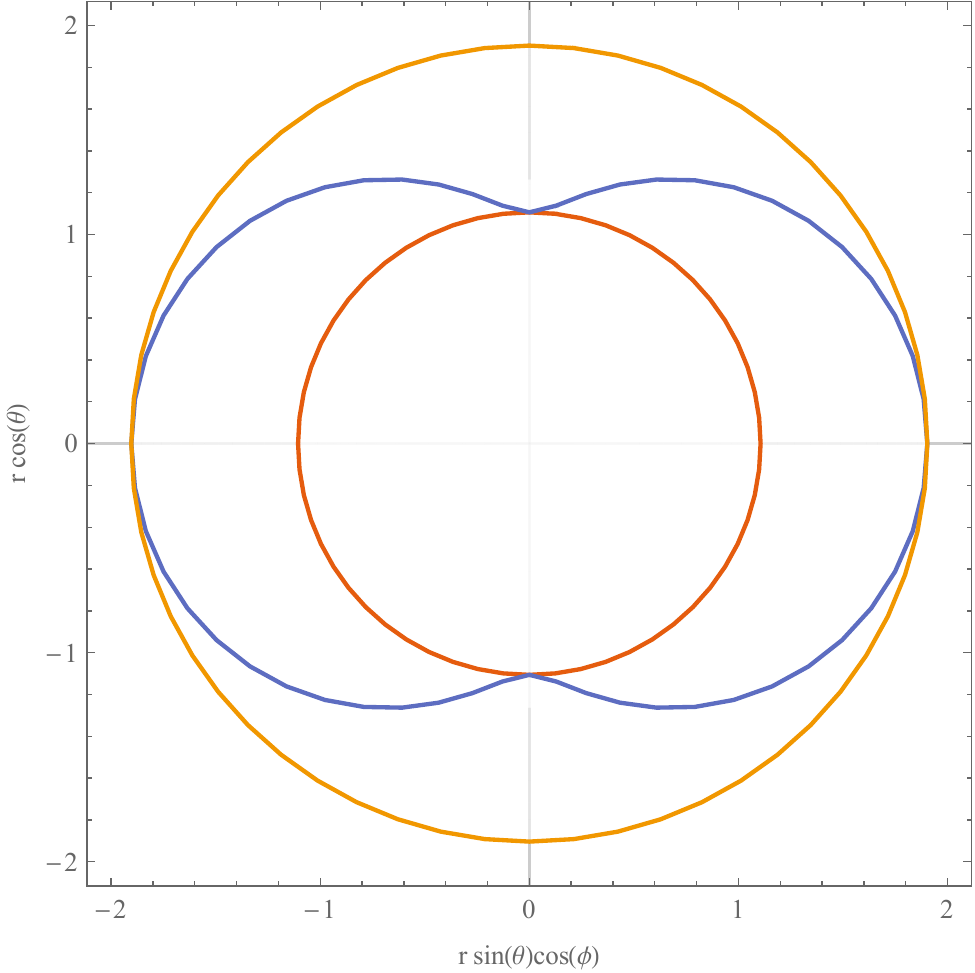}
		\includegraphics[width=8cm]{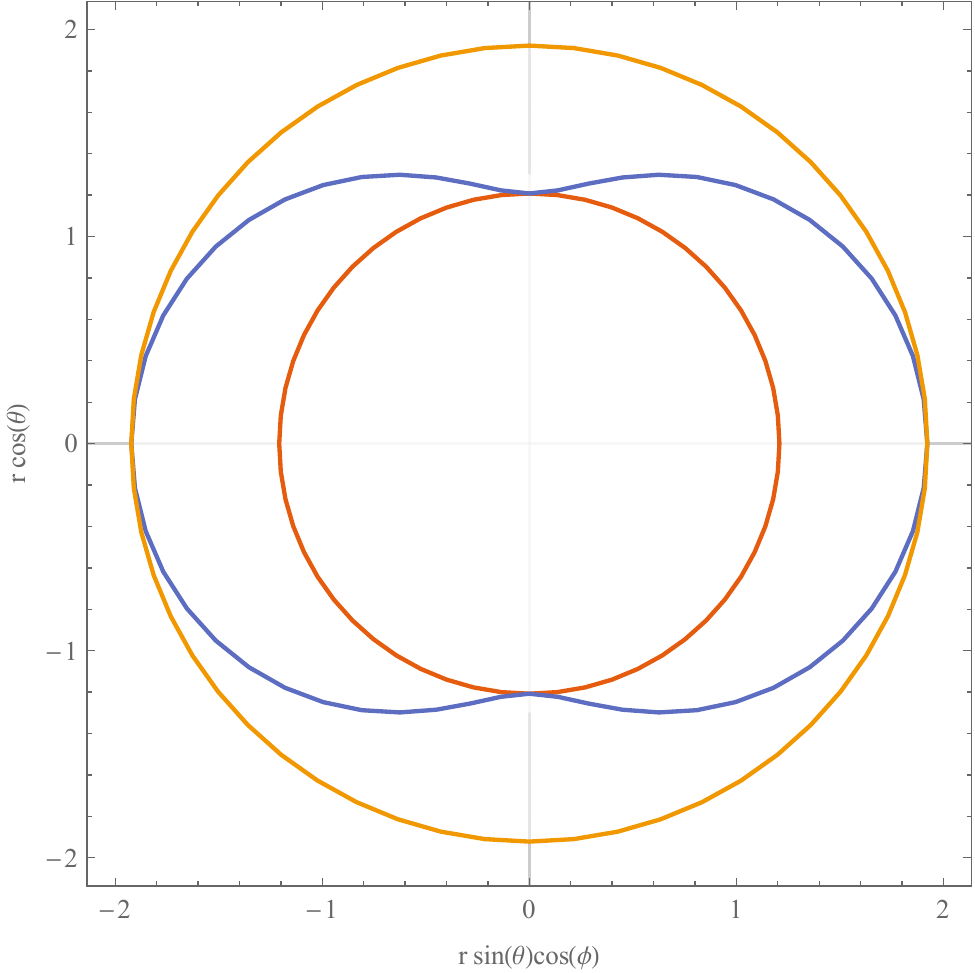}
		\includegraphics[width=8cm]{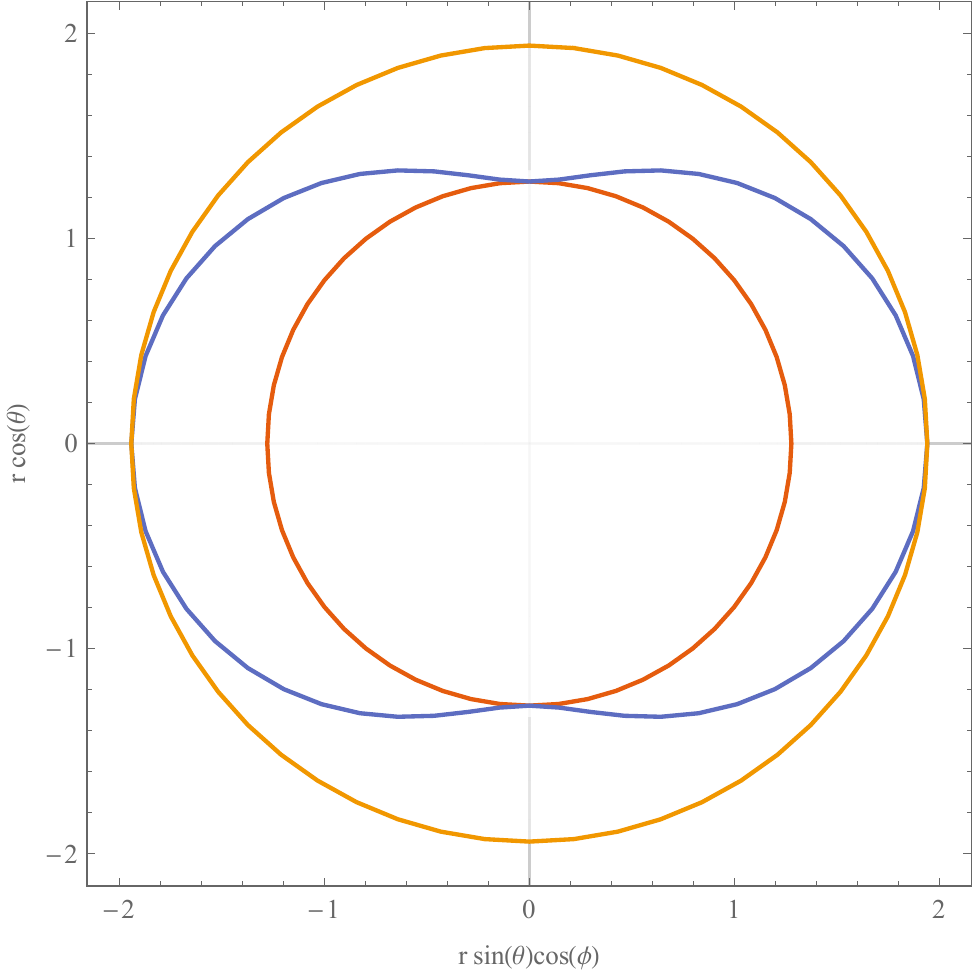}
		\includegraphics[width=8cm]{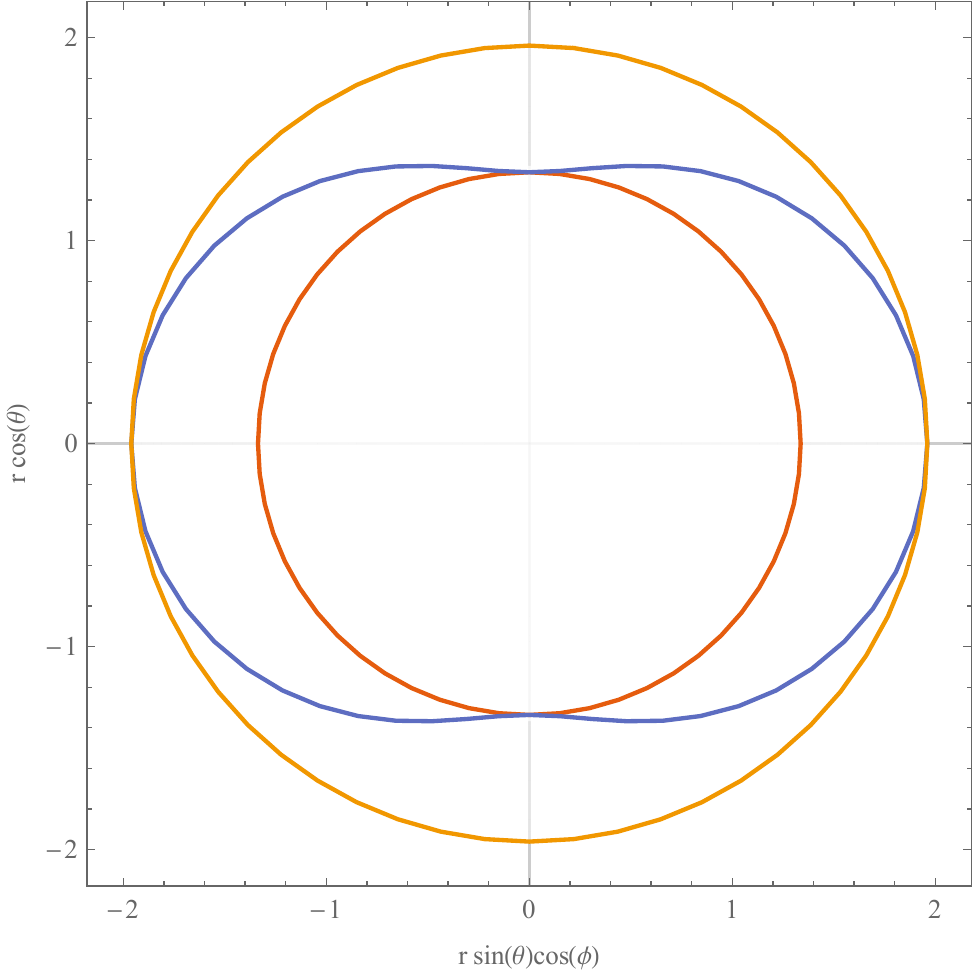}
		\caption{ Shapes of the outer horizons (red solid line) and outer ergo-sphere (blue solid line) in comparison with the outer event horizon in the static case (orange solid line) for {$M=1$} and $a=0.9$ with {different values of} $P$: upper and left panel: $P=0.025$; upper and right panel: $P=0.020$; bottom and left panel: $P=0.015$; and bottom and right panel: $P=0.01$.}
		\lb{bb1}
	\end{figure*}
	
	\subsection{Regularity of the spacetime}
	
	The regularity of a black hole solution can be {deduced} from analyzing the Kretschmann scalar $K$. This scalar, when tends to infinity, indicates the presence of a curvature singularity inside the black hole. One of
	{ the} remarkable features of {LQG} is that it could  provide  {a resolution} to the singularity of the spacetime. For spherically symmetric LQBH, it has been shown  that the singularity inside the Schwarschild black hole is cured after the quantum corrections of LQG are taken into account \cite{LQG_BH}. Therefore, it is natural to expect this is {also} the case for the rotating LQBH obtained in this paper. To see if the rotating LQBH is regular in the whole spacetime, let us study the Kretschmann scalar in the limit $r \to 0$ and $\theta=\pi/2$, 
	\bqn
	K&=&R^\mu_{\nu\sigma\rho}R^{\nu\sigma\rho}_\mu \nb\\
	&\simeq& \frac{64 M^4 P^4 [a_0^2 (1+P)^4+4 a^2 P^2 M^2	]}{a_0^8 (1+P)^{16}} + \mathcal{O}(r).
	\eqn
	It is {clear} that the rotating LQBH is {also} regular,  because of the presence of the minimal area $a_0=A_{\rm min}/8\pi$ arising from LQG.
	
	
\section{Null Geodesic and Circular photon orbits}
	
	In this section, we analyze the evolution of photons around the rotating LQBH. A photon follows a null geodesic in a given black hole spacetime. In order to find null geodesics around the black hole we can use the Hamilton-Jacobi equation given by,
	\bqn
	\frac{\partial S}{\partial \lambda}=-\frac{1}{2}g^{\mu\nu}\frac{\partial S}{\partial x^\mu}\frac{\partial S}{\partial x^\nu},
	\eqn
	where $\lambda$ is the affine parameter of the null geodesic and $S$ denotes the Jacobi action of the photon. The Jacobi action $S$ can be separated in the following form,
	\begin{equation}
	S=\frac{1}{2}m^2\lambda-Et+L\phi+S_r(r)+S_\theta(\theta),
	\end{equation}
	where $m$ denotes the mass of the particle moving in the black hole spacetime and for photon one has $m=0$. $E$ is the energy and $L$ represents the angular momentum of the photon in the direction of the rotation axis. The two functions $S_r(r)$ and $S_\theta(\theta)$ depend only on $r$ and $\theta$, respectively.
	
	Now substituting the Jacobi action into the Hamilton-Jacobi equation, we obtain
	\bqn
	S_r(r)&=&\int^r\frac{\sqrt{R(r)}}{\Delta}dr,\\
	S_\theta(\theta)&=&\int^\theta\sqrt{\Theta(\theta)}d\theta,
	\eqn
	\red{where
	\begin{eqnarray}
    R(r) &=&[X(r)E-aL]^2-\Delta(r)[\mathcal{K}+(L-aE)^2],\\
	\Theta(\theta)&=& \mathcal{K}+(a^2E^2-L^2\csc^2\theta)\cos^2\theta,
	\end{eqnarray}
	with $ \mathcal{K}$ denoting the Carter constant, where  $\Delta(r)$ is defined by Eq.(18), and $X(r)\equiv k+a^2$. }\footnote{\red{A typo in Eq.(34) is corrected, which subsequently propagates to several other equations, and the corresponding typos are also corrected, although our main results and conclusions remain the same. For more details, see \cite{erratum}}}
	
	Then, {the variations of the Jacobi action give rise} to the following four equations of motion for the evolution of the photon,
	\bqn
	\rho^2\frac{dt}{d\lambda} &=&a(L-aE\sin^2\theta) \nb\\
	&&+\frac{r^2+a^2}{\Delta}[(r^2+a^2)E -aL], \lb{YYY} \\
	\rho^2\frac{d\phi}{d\lambda} &=&\frac{L}{\sin^2\theta}-aE+\frac{a}{\Delta}[(R^2+a^2)E-aL],\\
	\rho^2\frac{dr}{d\lambda} &=&\sqrt{R(r)},\\
	\rho^2\frac{d\theta}{d\lambda}&=&\sqrt{\Theta(\theta)}.\lb{XXX}
	\eqn
	The motion of a photon is determined by the two impact parameters
	\bqn
	\xi=\frac{L}{E},\qquad \eta=\frac{\mathcal{K}}{E^2}.
	\eqn
	To determine the geometric shape of the shadow of the black hole, we need to find the critical circular orbit for the photon, which can be derived from the unstable condition
	\bqn
	R(r)=0,\qquad \frac{dR(r)}{dr}=0.\lb{ZZZ}
	\eqn
	The geometric shape of the shadow is determined by the allowed values of $\xi$ and $\eta$ that fulfill these conditions. In general, the shape of the shadow depends on {the rotation parameter $a$}.
	
	For a spherical symmetric LQBH, the shadow is a circular disk and it is described by the photon sphere. In this case, the above conditions (\ref{ZZZ}) reduce to
\red{	\bqn
	2g(r)h(r)\frac{d X(r)}{dr}-X(r) \frac{d}{dr}[g(r)h(r)]=0.
	\eqn
   Since the numerator of the above equation is a polynomial of order 7 in terms of $r$, we cannot solve it explicitly  to provide an analytical solution.
	However, the solution of the above equation determines the radius $r_{ps}$ of the photon sphere,
	\bqn\label{sum}
	\xi^2+\eta=\frac{h(r_\text{ps})}{f(r_\text{ps})}.
	\eqn    }
	
	For the rotating LQBH, solving the conditions (\ref{ZZZ}), one finds that, for the spherical motion of photons, the two impact parameters $\xi$ and $\eta$ assume the forms
\red{	\begin{align}  \label{AAA}
&\xi=\frac{X_\text{ps}\Delta'_\text{ps}-2\Delta_\text{ps}X'_\text{ps}}{a\Delta'_\text{ps}},\nb\\
&\eta=\frac{4a^2X'^2_\text{ps}\Delta_\text{ps}-\left[\left(X_\text{ps}-a^2\right)   \Delta'_\text{ps}-2X'_\text{ps}\Delta_\text{ps} \right]^2}{a^2\Delta'^2_\text{ps}},
	\end{align}
	where}
	
\red{	\bqn \label{BBB}
	X(r,M,a,P)&=&a^2+\frac{r^4+a_0}{(2 M P+r)^2},\nb\\
\Delta(r,M,a,P)&=&a^2+\frac{r^2 (2 M-r) \left(2 M P^2-r\right)}{(2 M P+r)^2}.
	\eqn}

	\red{The parameters $\xi$ and $\eta$ given by Eqs.~(\ref{AAA}) and~(\ref{BBB}) {reduce to} the expressions of {the} Kerr black hole when  $P=0, a_0=0$.} 
 As we shall show in the next section, the parameter $P$, a manifestation of the effects of LQG, tends to make the shadow smaller in size and more distorted in shape. Consequently, with the expressions of the two impact parameters given above, we will be able to find out the deviations from the Kerr spacetime. In the next section, we use these relations to discuss the shape and size of the shadow of the rotating LQBH.
	
	{Note that both expressions for ($\xi,\,\eta$) {diverge} as $a\to 0$, however, the expression $\xi^2+\eta$ has a finite value at $a=0$, as we shall see in the next section, and it coincides with the right-hand side of~\eqref{sum}.}

\section{Observables of Black hole shadow}
	
	In this section, we {aim} to construct the shape of the shadow of the rotating LQBH we obtained in Sec. II.  In general, the  {photons} emitted by a light source will be deflected when  {they pass} by the black hole because of the gravitational lensing effects. Some of the  {photons} can reach the distant observer after being deflected by the black hole, and some of them directly fall into the black hole. The photons that cannot escape from the black hole form the shadow of the black hole in the observer's sky. The border of the shadow defines the apparent shape of the black hole. To study the shadow, we adopt the celestial coordinates:
	\begin{eqnarray}
	\alpha&=&{\lim_{r_0\to \infty}\left( \left.-r_0^2 \sin{\theta_0}\frac{d \phi}{dr}\right|_{\theta\rightarrow\theta_0} \right)=-\xi \csc{\theta_0}},\\
	\beta&=&\lim_{r_0\to \infty} \left(\left.r_0^2\frac{d\theta}{dr}\right|_{\theta\rightarrow\theta_0}\right) \nb\\
	&=& \pm\sqrt{\eta+a^2 \cos^2\theta_0-\xi^2 \cot^2\theta_0}\;,
	\end{eqnarray}
	where we have used Eqs.~(\ref{YYY})-(\ref{XXX}), $r_0$ denotes the distance between the observer and the black hole and $\theta_0$ represents the inclination angle between the line of sight of the observer and the rotational axis of the rotating LQBH. In the special case where the observer is on the equatorial plane of the black hole with the inclination angle $\theta_0=\pi/2$, one obtains
	\begin{eqnarray}
	\alpha&=&-\xi,\\
	\beta&=&\pm\sqrt{\eta}.
	\end{eqnarray}
	It is easy to {show} that the two celestial coordinates satisfy
	\begin{equation}
	\alpha^2+\beta^2=\xi^2+\eta.
	\end{equation}
	Using Eqs.~(\ref{AAA}) and~(\ref{BBB}) we obtain 
\red{
\begin{align}\lb{contour}
\alpha^2+\beta^2=\frac{4 X'\Delta   \left(X'-\Delta '\right)}{\Delta '^2}+2 X-a^2.
		\end{align}
}

	Now we can calculate the observables and plot the shape of the shadow for the rotating LQBH by using Eq. (\ref{contour}). According to (\ref{contour}), it {can be seen} that the shape and size of the shadow for the rotating LQBH depend on its specific angular momentum $a$, the inclination angle [which was set $=\pi/2$ in~\eqref{contour}], and the polymeric function $P$. We shall plot $\alpha$ v.s. $\beta$ to display the shape and size of the shadow for various values of $a$ and $P$ at different inclination angles. {In the following, {to make comparisons,} we consider the non-rotating LQBH and the rotating LQBH, respectively}.

\subsection{Non-rotating LQBH}
	
	\begin{figure}
		\centering
		\includegraphics[width=8cm]{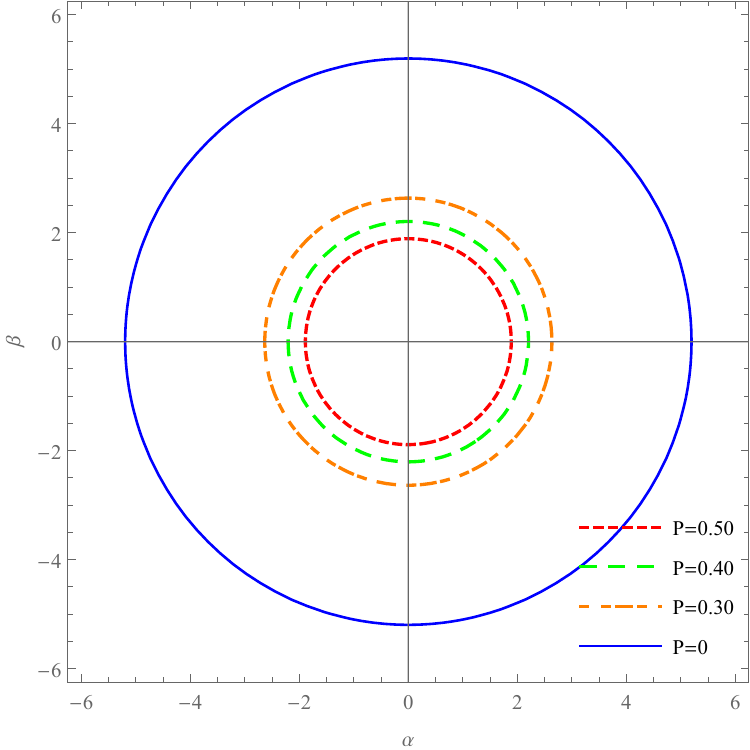}
		\caption{ Shadow of {the} non-rotating LQBH for different values of the polymeric function $P$. The blue solid circle corresponds to the shadow of the Schwarzschild black hole. \red{Here we have chosen $a_0=0.$}}
		\lb{a1}
	\end{figure}

	\begin{figure*}
		\centering
		\includegraphics[width=8.1 cm]{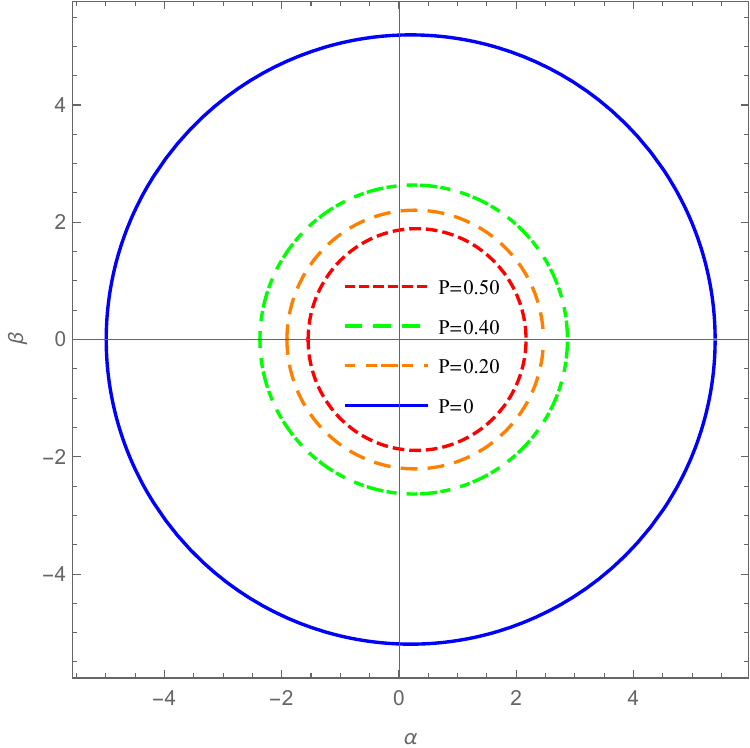}
		\includegraphics[width=7.94 cm]{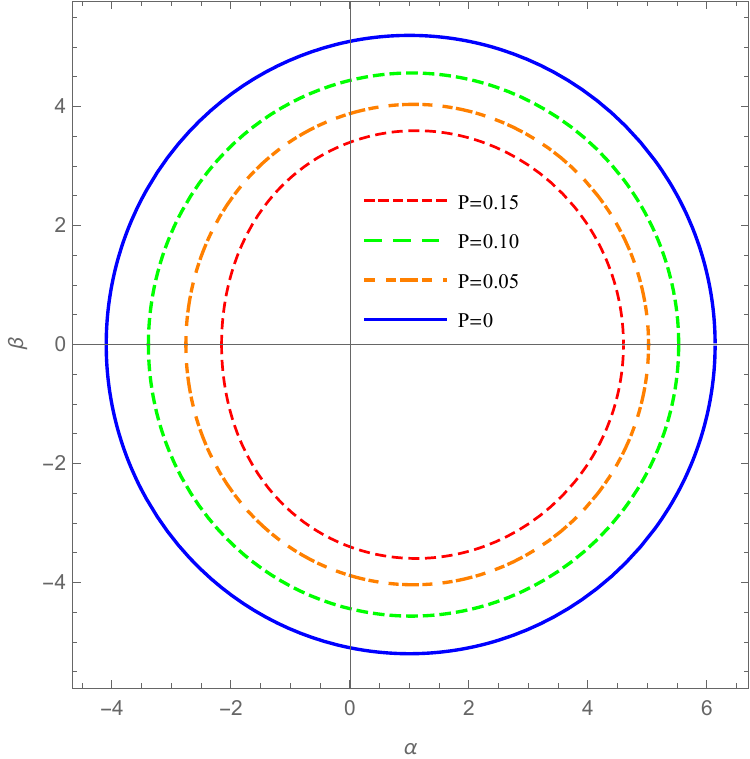}
		\includegraphics[width=8.2cm]{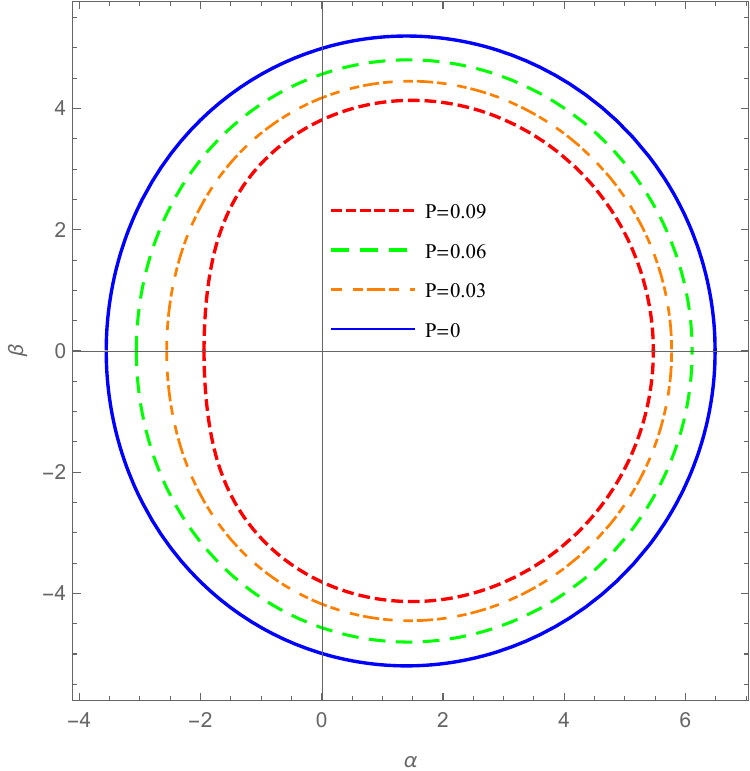}
		\includegraphics[width=7.9cm]{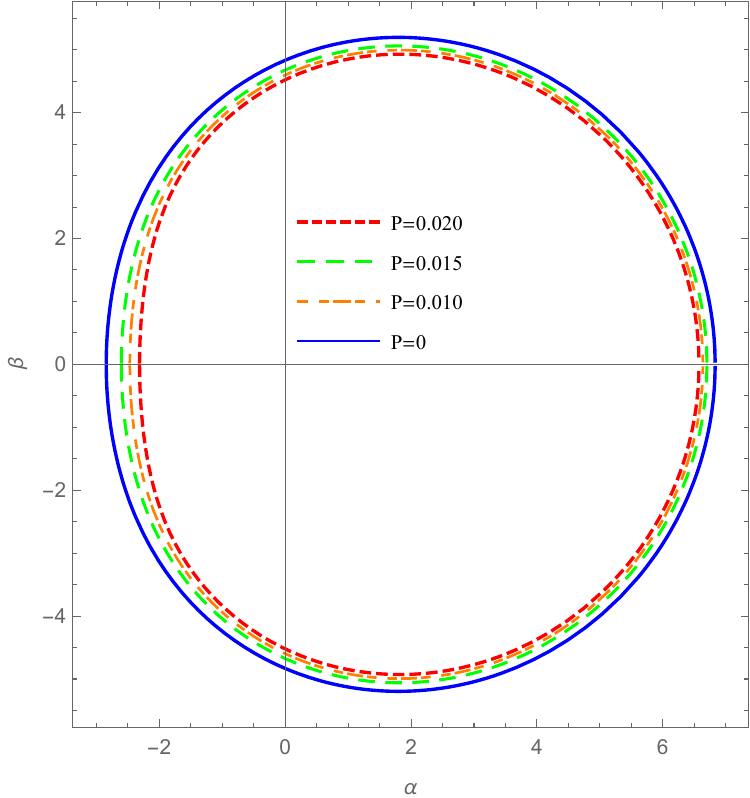}
		\caption{ Shadow of {the} rotating LQBH for different values of $a$, {$\theta_0$}, and $P$. The blue solid curve corresponds to the shadow of the Kerr black hole. Upper and left panel: $r_{+}/M=2$, {$\theta_0 = 17^o$}, $a=0.1$; upper and right panel: $r_{+}/M=2$, {$\theta_0 = 90^o$}, $a=0.5$; bottom and left panel: $r_{+}/M=2$, {$\theta_0 = 90^o$}, $a=0.8$; {and bottom} and right panel: $r_{+}/M=2$, {$\theta_0 = 90^o$}, $a=0.9$. \red{For all plots we have chosen $a_0=0$.}}
		\lb{bcd}
	\end{figure*}
	
	We first consider the  non-rotating case. For this case,  one can easily find that the shape of the shadow is a standard {disk}, while its radius is closely dependent on the mass $M$ of the black hole  and the polymeric function $P$. The observable $R_s$, which is the apparent size of the shadow for the spherical symmetric LQBH, can be calculated via
\red{
	\bqn
	R_s=\sqrt{\alpha^2+\beta^2}=\sqrt{\frac{h(r_{ps})}{f(r_{ps})}}.
	\eqn  }
	The shape of the shadow for the spherical symmetric LQBH ($a=0$) is plotted in Fig.~\ref{a1} for different values of the polymeric function $P$.  It is   {obvious} that the size of the shadow region decreases with increasing $P$.
	
	The polymeric function $P$ arising from {LQG} is expected to be constrained by the measurement of the angular diameter of the shadow. For the central supermassive black hole in M87*, depending on the distance $D$ between the black hole and the Earth, the angular {diameter} of the shadow seen by the observer is given by,
	\bqn\label{Rs}
	\theta_s = 2 R_s /D,
	\eqn
	which is $\theta_s = (42 \pm 3) {\rm \mu as}$ as measured from the first image of the black hole by EHT \cite{m87}. However, {in the} static LQBH case, the polymeric function $P$ and the black hole mass $M$ are degenerate. They both affect the apparent size of the shadow,  and the nonzero polymeric function $P$ may increase the mass of the supermassive black hole in M87*. In this case, the constraint on $P$ is only possible if the mass can be measured precisely in other independent observations. Therefore, if the mass of the black hole can be measured independently from other observations, it will break the parameter degeneracy and possibly lead to interesting bounds on the parameter space $(M, P)$.
	
	\red{Conversely, one may try to constrain $M$ under the case $a_0=0$.} A Taylor expansion of the right-hand side of~\eqref{Rs} about $P=0$ yields
\red{	\begin{equation}\label{Rs1}
	\theta _s=\frac{2 (27 M-72 M P+\cdots)}{3 \sqrt{3} D}, \;\text{ if }\; P\ll 1,
	\end{equation}
	and
\begin{equation}\label{Rs2}
\theta _s=\frac{2 (27 M-36 M P+136 M P^2+\cdots)}{3 \sqrt{3} D}, \;\text{ if }\; P< 1.
\end{equation}
Equation~\eqref{Rs1} yields
\begin{equation}\label{Rs3}
P=\frac{3}{4}-\frac{D \theta _s}{16 \sqrt{3} M}\gtrsim 0\Rightarrow M\gtrsim \frac{D\theta_s}{6\sqrt{3}},
\end{equation}
and Eq.~\eqref{Rs2} yields
\begin{multline}\label{Rs4}
P=\frac{18 M\pm \sqrt{3} \sqrt{M (-198 M+17 \sqrt{3} D \theta _s)}}{68 M}\\
\Rightarrow M\leq \frac{1.5 D\theta_s}{6\sqrt{3}}.
\end{multline}
From~\eqref{Rs3} and~\eqref{Rs4} we obtain the following constraint on $M$ under the condition $a_0=0$,  resulting from the effects of LQG in the spherical case of black holes,
\begin{equation}\label{Rs5}
\frac{1.5 D\theta_s}{6\sqrt{3}}\geq M \gtrsim \frac{D\theta_s}{6\sqrt{3}}.
	\end{equation}}
	
	
\subsection{Rotating LQBH}
	
	 { In the rotating case ($a \neq0 $), the shadow in the parameter  space $(\alpha, \beta)$ will be of a deformed circle due to the} dragging effects. In Fig.~\ref{bcd}, we display the contour of $\alpha$ and $\beta$ in Eq.~(\ref{contour}) {delineating} the shadow of the rotating LQBH for different  {values of} the angular  momentum $a$,  inclination angle $\theta_0$, and polymeric function $P$. The shadow corresponds to the region inside each closed curve. These figures reveal the interesting property that the shape of the black hole shadow changes with respect to the polymeric function $P$ for fixed $a$ and  $\theta_0$. For a fixed angular momentum, the shape of the {shadow} gets more deformed from {circularity} as $P$ increases. {For lower values of the angular momentum $a$, the shape is almost circular, while the deviation becomes significant when the rotation parameter approaches a large value, i.e., $a=0.9$. For the case $P=0$, our result coincides with that for the Kerr black hole.}
	
	Now let us turn to consider the actual size and distortion of the shadow for the rotating LQBH. For this purpose, we define two observables that characterize the black hole shadow, namely, $R_s$ and $\delta_s$, where the parameters $R_s$ and $\delta_s$ correspond to the actual size of the shadow and distortion in the shape of the shadow, respectively.  {We approximate the shadow periphery by a reference circle that coincides at the top, bottom, and extreme right edges with the shadow.}
	{The size of the shadow $R_s$ is the radius of a reference circle passing by the three points: the top point $(\alpha_t, \beta_t)$, the bottom one $(\alpha_b, \beta_b)$, and the most right one $(\alpha_r, 0)$ of the shadow.
The points $(\alpha_p, 0)$ and $(\tilde{\alpha}_p, 0)$ are points where the circle of the shadow and the reference circle cross the horizontal axis at the opposite side of $(\alpha_r, 0)$, respectively. The radius $R_s$ gives the approximate size of the shadow, and $\delta_s$ measures its deformation with respect to the reference circle. Performing a simple algebra calculation, one determines the radius  {of} the reference circle by
	\bqn
	R_s=\frac{(\alpha_t-\alpha_r)^2+\beta_t^2}{2(\alpha_r-\alpha_t)},
	\eqn
	where we have used the relations $\alpha_b=\alpha_t$ and $\beta_b=-\beta_t$ because of the symmetry of the shadow. Now,
	\bqn
	\delta_s=\frac{D_s}{R_s}=\frac{|\alpha_p-\tilde{\alpha}_p|}{R_s},
	\eqn
	where $D_s$ is the distance from the most left point $(\alpha_p, 0)$ of the shadow to the most right point of the reference circle $(\tilde \alpha_p, 0)$. Considering the relation $\tilde{\alpha}_p=\alpha_r-2R_s$, we obtain
	\bqn
	\delta_s=2-\frac{d_s}{R_s},
	\eqn
	where $d_s=\alpha_r-\alpha_p$ is the diameter of the shadow along the $\alpha$-axis. For the shadow cast by the spherically symmetric LQBH, the outline of the shadow will coincide with the reference circle, so there will be no distortion, i.e. $\delta_s=0$. We numerically calculate these observables and the results are presented in Fig.~\ref{abcd}. It is evident that the presence of the polymeric function $P$ from  {LQG} has a profound influence on the apparent shape and size of the shadow. It is shown clearly in Fig.~\ref{abcd} that, {as the polymeric function $P$ increases, the shadow size  gradually decreases, while  its distortion gradually increases,   for  fixed  $\theta_0$ and  $a$}.
	
	
	\begin{figure*}
		\centering
		\includegraphics[width=7.9cm]{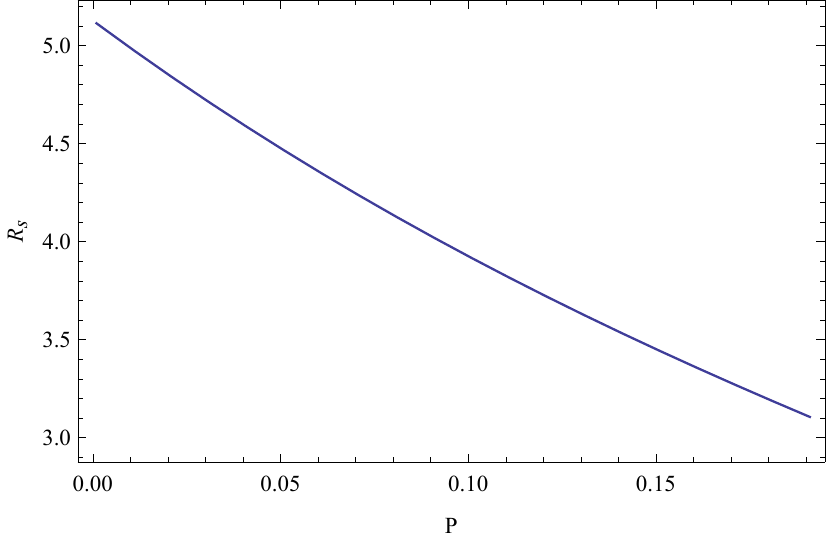}
		\includegraphics[width=7.9cm]{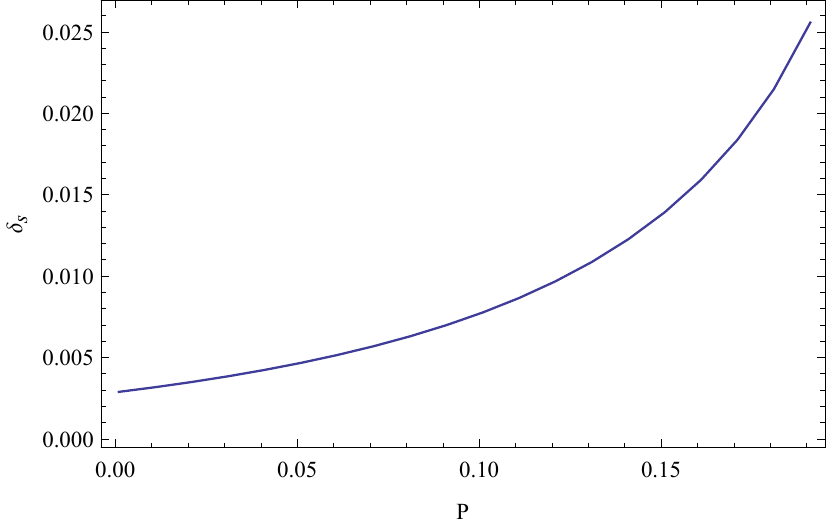}
		\caption{ Behavior of the observables $R_s$ (left panel) and $\delta_s$ (right panel) as function of the parameter $P$. \red{In these plots we set $M=1$, $a=0.5$ and $a_0=0$.}}
		\lb{abcd}
	\end{figure*}
	
	{Here we  {would like to go further and see how the effects of the polymeric function $P$ are tied to the recent observation of the black hole shadow in M87* by  EHT Collaboration.} The observation indicates that the shadow is nearly circular and the deviation from circularity is less than $10\%$.  For  {the} Kerr spacetime, it is interesting to note that this observable is independent of the measurement of the mass of the black hole, and is very sensitive to the angular momentum of the black hole. For  {the} rotating LQBH, since the polymeric function $P$ increases the distortion of the shadow, one expects that it could lead to some significant contribution in the deviation from circularity. In order to see how the polymeric function $P$ can affect the deviation from the circularity of the shadow, let us consider the average radius of the shadow,
	\bqn
	\bar R = \frac{1}{2\pi} \int^{2\pi}_0 \sqrt{(\alpha- \alpha_c)^2 + \beta^2}~d \vartheta,
	\eqn
	where $(\alpha_c, 0)$ denotes the geometric center of the shadow and $\vartheta$ determines the angle along the shadow boundary from the $\alpha$-axis,
	\bqn
	\vartheta = \arctan{\left(\frac{\beta}{\alpha - \alpha_c}\right)}.
	\eqn
	Then the deviation from the circularity of the shadow is defined as
	\bqn
	\Delta C =  2 \sqrt{ \frac{1}{2\pi} \int_0^{2\pi} \Big[\sqrt{(\alpha- \alpha_c)^2 + \beta^2} - \bar R \Big]^2 d \vartheta}.
	\eqn
	
	\begin{figure*}
		\centering
		\includegraphics[width=8.2cm]{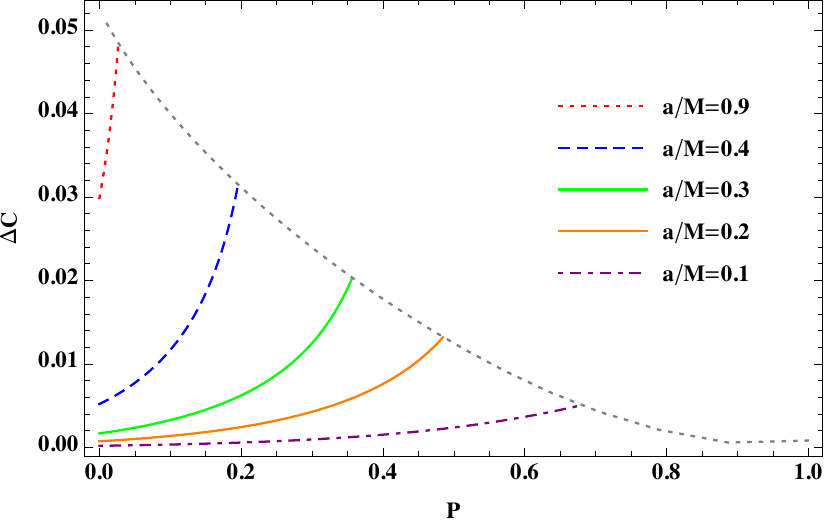}
		\includegraphics[width=8.2cm]{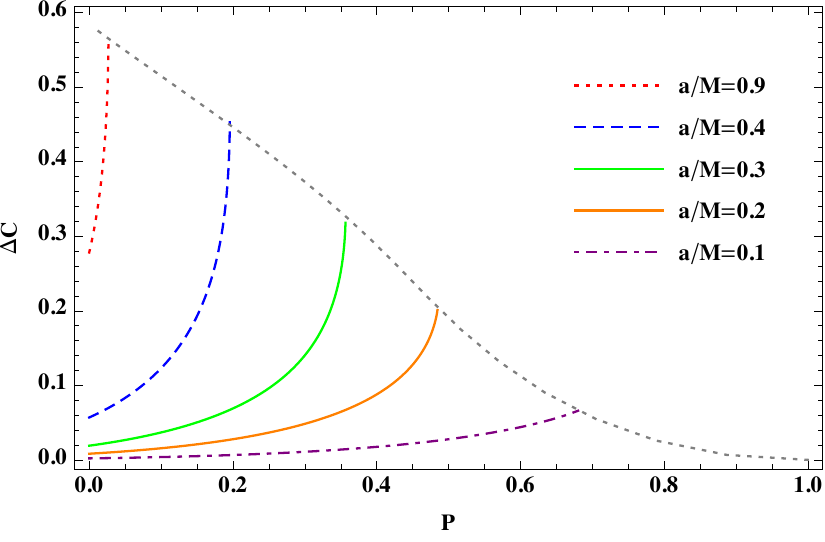}
		\caption{The deviation from circularity of the shadow {versus the polymeric function $P$} for different values of $a$ and  $\theta_0$. Left panel: $\theta_0 =17^o $;  {and} right panel: $\theta_0 =90^o $. The gray dotted curve in each panel denotes the maximum value of the polymeric function $P$ for different values of the spin $a$.}
		\lb{DC}
	\end{figure*}
	
	We numerically calculate the circularity of the shadow $\Delta C$ and plot it as  {a} function of the polymeric function $P$ for different values of  {the spin parameter} $a$ and inclination angle $\theta_0$ in Fig.~\ref{DC}. The left panel corresponds to $\theta_0=17^o$. The cases for the spin values $a=0.9$, $0.4$, $0.3$, $0.2$, and $0.1$ are, respectively, described by the red dotted line, blue dashed line, green solid line, orange solid line, and purple dotted dashing line. The gray dotted curve in each panel denotes the maximum value of the polymeric function $P$ for different values of  $a$. Therefore, the region above the gray dotted curve is the forbidden region of the parameter space $(P,\, \Delta C)$. For both panels, one can easily find that $\Delta C$ increases with increasing $P$ for  {fixed} $a$. Similarly, for a given $P$, $\Delta C$ also increases with increasing $a$. Comparing the two panels, it is interesting to note that the inclination angle has a significant influence on the circularity of the shadow $\Delta C$, which significantly increases as the inclination angle increases from $\theta_0 = 17^o$ to $\theta_0 = 90^o$. Considering the inclination angle for  {the} black hole shadow of M87* is about $\theta_0=17^o$, it is evident to infer from the left panel of Fig.~\ref{DC} that all the allowed values of $(a, P)$ are consistent with the observational data, which implies  {that} one will not be able to get any meaningful bounds on $(a,\, P)$. One expects that future precise measurement of the deviation from circularity for the black hole shadow can have capacity to give interesting bounds on  $(a, \,P)$ for a given inclination angle.
	
%

\section{Connection between the shadow radius and QNMs }

In a seminal paper by Cardoso et al. \cite{cardoso},  it was argued that in the eikonal limit, the real part of the QNMs is related to the angular velocity of the last circular, null geodesic, while the imaginary part of
{the} QNMs is related to the Lyapunov exponent that determines the instability time scale of the orbit,
\begin{equation}
	\omega_{QNM}=\Omega_c l -i \left(n+\frac{1}{2}\right)|\lambda|.
\end{equation}
Furthermore,  this important result is expected to be valid not only for static spacetimes but also for rotating black holes. Later on, Stefanov et al. \cite{Stefanov:2010xz} pointed out a connection between black-hole QNMs in the eikonal limit and strong lensing. Very recently, {Jusufi \cite{Jusufi:2019ltj}} pointed out that the real part of the QNMs and {the} shadow radius are related by the following relation:
\begin{equation}\label{k1}
	\omega_{\Re} = \lim_{l \gg 1} \frac{l}{R_S},
\end{equation}
which is precise only in the eikonal limit having large values of $l$. Hence we can write
\begin{equation}
	\omega_{QNM}=\lim_{l \gg 1} \frac{l}{R_S} -i \left(n+\frac{1}{2}\right)|\lambda|.
\end{equation}

In other words, instead of the angular velocity, it is more convenient to express the real part of QNMs in terms of {the} black hole shadow radius. This close connection can be understood from the fact that the gravitational waves can be treated as massless particles propagating along the last null unstable {orbit  out} to infinity. At this point, we note that the correspondence is not guaranteed for gravitational fields, as the link between the null geodesics and
QNMs is shown to be violated in the context of the Einstein-Lovelock theory even in the eikonal limit \cite{Konoplya:2017wot}. Although the relation~\eqref{k1} is not accurate for small $l$, as we are going to see, it can provide important {information about} the effect of the quantity $P$ on the shadow radius once we have calculated the real part of QNMs or vice verse. In what follows, we are going to elaborate the effect of $P$ on the QNMs in the context of the effective LQG. Toward this purpose,  let us start from the corrected Schwarzschild metric (\ref{1}) and introduce the following coordinate transformation
\begin{equation}
	dr_{\star}=\frac{dr}{\sqrt{f(r) g(r)}},
\end{equation}
{which yields} the following metric
\begin{equation}
	ds^2=f(r_{\star})\left(dt^2-dr^2_{\star}\right)-h(r_{\star})\left(d\theta^2+\sin^2\theta \,d\phi^2\right).
\end{equation}
\begin{figure*}
	\centering
		\includegraphics[width=8.2 cm]{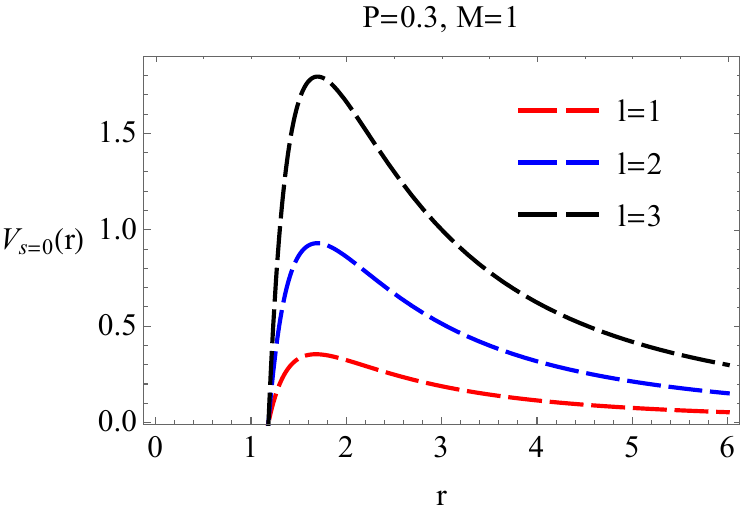}
		\includegraphics[width=8.2 cm]{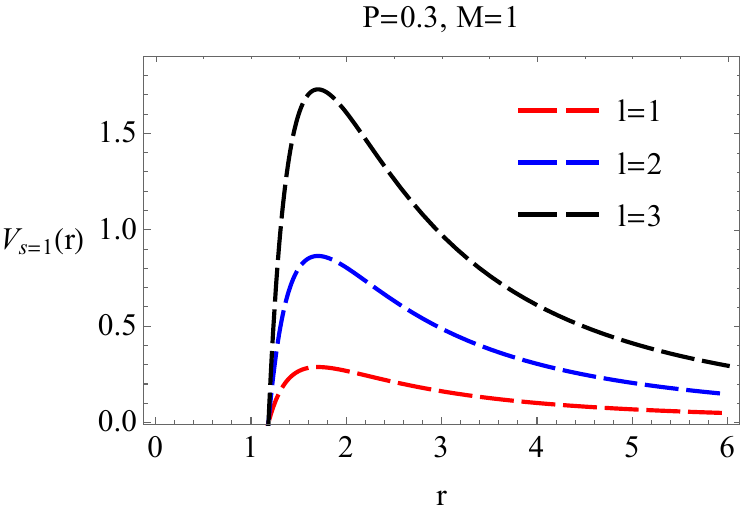}
		\caption{Left panel: {The} effective potential of the scalar field perturbation $V_{s=0}$ for different values of $l$. Right panel: The effective potential of the electromagnetic field perturbation $V_{s=1}$ for different values of $l$. For both plots we have chosen $a_0=0.01$. Changing the values of $l$ changes the height of the potential barrier. In the case of {the} scalar perturbations the height of the potential barrier is {slightly} higher compared to the case of  {the} electromagnetic perturbations.}
	\label{fig8}
\end{figure*}
Using the formalism developed in \cite{Boonserm:2013dua}, the Regge-Wheeler equation can be written as follows
\begin{equation}
	\partial_{\star}^2\Phi+\left(\omega^2-V_s(r)\right)\Phi=0,
\end{equation}
where the effective potential is given by
\begin{multline}
	V_s(r)=f(r) \left[\frac{l(l+1)}{h(r)}+\frac{s(s-1)(g(r)-1)}{h(r)} \right]\\+ (1-s)\frac{\partial^2_{\star} \sqrt{h(r)}}{\sqrt{h(r)}}
\end{multline}
with $s=0$ denoting the case for {the} scalar field while $s=1$ representing the case for {the} electromagnetic field. Having the expression for the effective potential one can use the WKB approach  to compute the {QNM} frequencies. As it is well known, the WKB method is basically used to solve the problem of {waves} scattering near the peak of the potential barrier. This method was
used by Schutz and Will \cite{Schutz}, then developed to the third order by Iyer and Will \cite{{Iyer}} with the QNM frequencies given by

\begin{figure*}
	\centering
	\includegraphics[width=8cm]{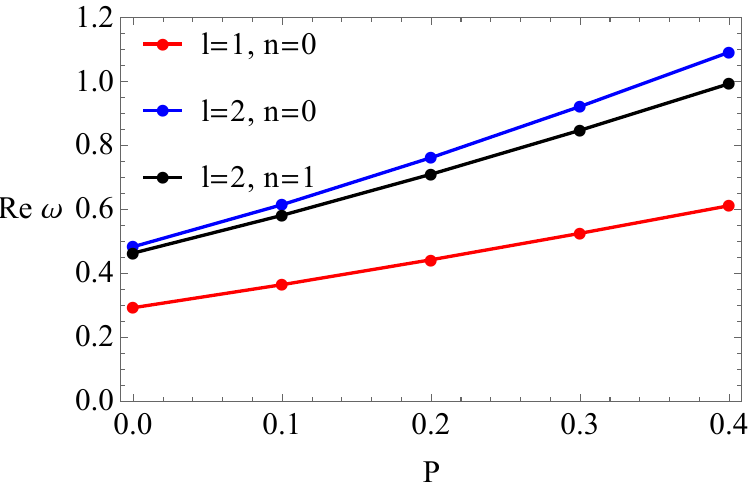}
	\includegraphics[width=8cm]{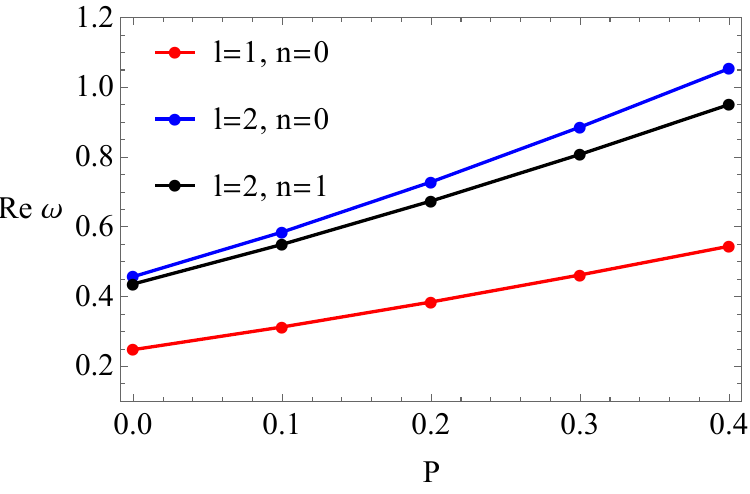}
	\caption{Left panel: Plots of the real part of {the} QNM frequencies versus the polymetric function $P$ for the scalar test field. Right panel: Plots of the real part of {the}
	QNM frequencies versus the polymetric function $P$ for the electromagnetic field. In both  cases we have chosen $M=1$ and $a_0=0.01$.}
	\lb{re1}
\end{figure*}
\begin{equation}
	\omega^2=\left[V_0+\sqrt{-2\,V_0''}\Lambda_2\right]-i\left(n+\frac{1}{2}\right)\sqrt{-2\,V_0''}(1+\Lambda_3),
\end{equation}
where
\begin{eqnarray}\notag
	\Lambda_2 &=& \frac{1}{\sqrt{-2\,V_0''}}\Big\{\frac{1}{8}\Big(\frac{V_0^{(4)}}{V_0''}\Big)
	\Big(\frac{1}{4}+\alpha^2\Big)\\
	&-&\frac{1}{288}\Big(\frac{V_0^{(3)}}{V_0''}\Big)^2(7+60\alpha^2)\Big\},
	\nonumber\\\notag
	\Lambda_3 &=& \frac{1}{\sqrt{-2\,V_0''}}\Big\{\frac{5}{6912}\Big(\frac{V_0^{(3)}}{V_0''}\Big)^4
	\Big(77+188\alpha^2\Big)\\\notag
	&-&\frac{1}{384}\Big(\frac{V_0'''^2V_0^{(4)}}{V_0''^3}\Big)(51+100\alpha^2)
	\nonumber\\\notag
	&+&\frac{1}{2304}\Big(\frac{V_0^{(4)}}{V_0''}\Big)^2(67+68\alpha^2)\\
	&-&\frac{1}{288}\Big(\frac{V_0'''V_0^{(5)}}{V_0''^2}\Big)(19+28\alpha^2)\nonumber\\
	&-&\frac{1}{288}\Big(\frac{V_0^{(6)}}{V_0''}\Big)(5+4\alpha^2)\Big\},
\end{eqnarray}
and
\begin{eqnarray}
	\alpha=n+\frac{1}{2},\;\;V_0^{(m)}=\frac{d^mV}{dr_*^m}\Big|_{r_{\star}}.
\end{eqnarray}
In the present paper, we are going to use the sixth order WKB approximation developed by Konoplya \cite{KonoplyaWKB}
\begin{equation}
	i\frac{\omega_n^2-V_0}{\sqrt{-2\,V_0''}}-\sum_{i=2}^{6}\Lambda_i=n+\frac{1}{2}
\end{equation}
where the constants $\Lambda_4,\;\Lambda_5,\;\Lambda_6$ can be found in  \cite{KonoplyaWKB}.  Note that $V_0$ represent the height of the barrier and $V_0''$ stands for the second derivative with respect to the tortoise coordinate of the potential at the maximum. The corresponding potentials for {the} scalar and electromagnetic fields are given in Fig.~\ref{fig8}.
The corrections depend on the {values} of the potential and higher derivatives of it at the maximum. We have presented our {QNM} results for the scalar and electromagnetic perturbations in Table I and Table II, respectively.

\begin{figure*}
	\centering
	\includegraphics[width=8cm]{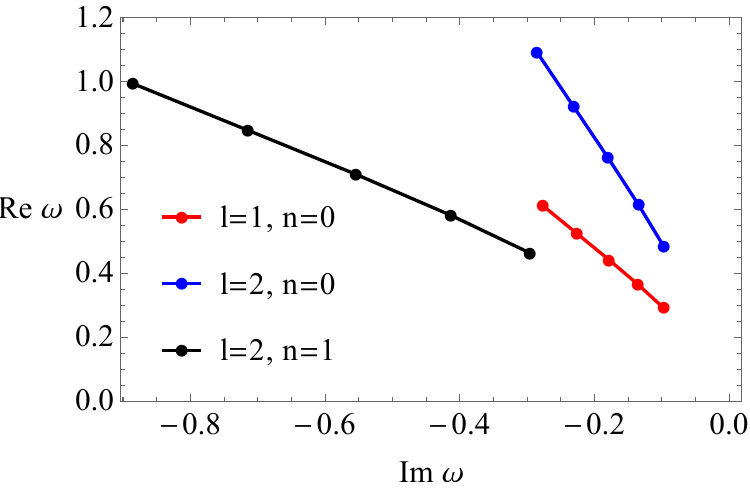}
	\includegraphics[width=8cm]{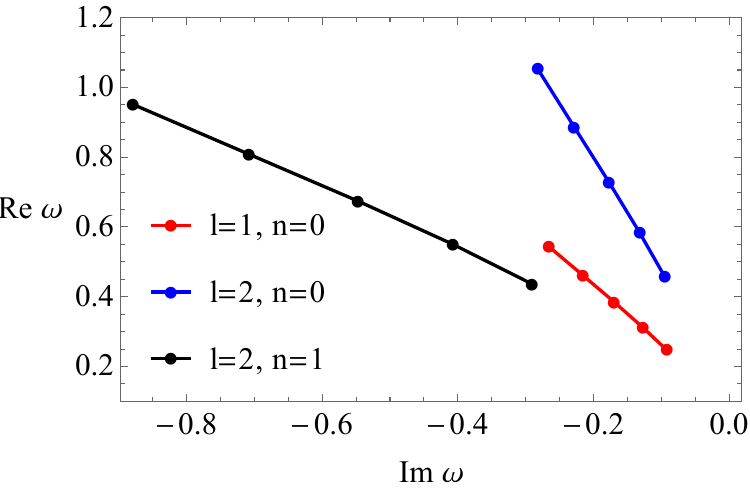}
	\caption{Left panel: {Plots of the real part of {the} QNM frequencies versus its imaginary part for the scalar test field. Right panel: Plots of the real part of {the} QNM frequencies versus its imaginary part for the electromagnetic field}. We have chosen $M=1$ and $a_0=0.01$.}\lb{re2}
\end{figure*}

\begin{table}[tbp]
	\begin{tabular}{|l|l|l|l|l|}
		\hline
		\multicolumn{1}{|c|}{ spin 0 } &  \multicolumn{1}{c|}{  $l=1, n=0$ } & \multicolumn{1}{c|}{  $l=2, n=0$ } & \multicolumn{1}{c|}{ $l=2, n=1$ }\\\hline
		$P$ & $\omega \,(WKB)$ &  $\omega \,(WKB)$ &  $\omega \,(WKB)$  \\ \hline
		0 & 0.2929-0.0978 i & 0.4836-0.0968 i & 0.4638-0.2956 i  \\
		0.1 & 0.3650-0.1350 i & 0.6153-0.1347 i & 0.5820-0.4133 i    \\
		0.2 & 0.4429-0.1785 i & 0.7624-0.1800 i & 0.7105-0.5546 i  \\
		0.3 & 0.5256-0.2265 i & 0.9224-0.2311 i & 0.8480-0.7147 i  \\
		0.4 & 0.6120-0.2766 i & 1.0920-0.2853 i & 0.9935-0.8848 i \\\hline
	\end{tabular}
	\caption{The real and imaginary {parts} of QNMs of the scalar field with {$M=1$,  $a_0=0.01$,} and different values of $P$. }\label{tab1}
\end{table}
From Table \ref{tab1} and Table \ref{tab2} (but see also Fig.~\ref{re1}), we see that by increasing $P$, the real part of QNMs and the imaginary part of QNMs {increase} in {their absolute values}. Moreover, we see that for the scalar field perturbations the values of the real part/imaginary part of QNMs in absolute {values} are higher compared to the electromagnetic field perturbations. This means that the scalar field perturbations decay more rapidly compared to the electromagnetic field perturbations. In general, however, the QNMs for the LQBH deviate from the vacuum Schwarzschild black hole due to the quantum effects,  and for any $P > 0$, the field perturbations induced by quantum effects decay more rapidly compared to {the} Schwarzschild vacuum black hole (see Fig.~\ref{re2}). Once we have computed the real part of QNMs and shown that $\omega_{\Re }$ increases with $P$, we can make use of the inverse relation between  $\omega_{\Re }$ and the shadow radius $R_S$
\begin{equation}\label{inv}
	\omega_{\Re }(P) \propto \frac{1}{R_S(P)},
\end{equation}
to show that the shadow radius decreases with increasing $P$, {as can be seen from} Figs.~\ref{a1} and~\ref{bcd} from {the} geodesic method. It's quite amazing that we can deduce this information directly from the inverse relation between the real part of QNMs and the shadow radius~\eqref{k1} even in {the} case of small multipoles $l$, although the relation~\eqref{k1} is precise only in the eikonal regime,  $l>>1$. Note that the gravitational perturbations in the spacetime of metric~\eqref{1} has been investigated in Refs. \cite{Cruz:2015bcj} and \cite{Moulin:2019ekf}. {In Ref. \cite{Moulin:2019ekf} the metric of the non-rotating loop quantum black hole was expressed in terms of the mass parameter $m$ which is related to the ADM mass $M$ by $M=(1+P)^2m$. However, some care must be taken due to the fact that the physical mass measured by an observer located at spatial infinity is the ADM mass $M$ and not the mass parameter $m$. Taking into consideration this fact, we have expressed in this present work, the metrics of the non-rotating and rotating loop quantum black holes in terms of the ADM mass to study the shadow and the QNMs. From the shadow plots, Fig.~\ref{Rs_QNMS}, we see that the shadow radius decreases meaning that the real part of QNMs must increase, as shown in Fig.~\ref{re1}, due to the inverse relation between them~\eqref{inv}. We have verified this fact using the WKB method to compute the QNMs. This seems to be not the case in Ref.~\cite{Moulin:2019ekf} where a non-monotonic behavior of the real part of QNMs in terms of $\delta$, with $\gamma$ held constant, has been obtained. When $\gamma$ is held constant $P$ becomes a monotonically increasing function of $\delta$. This means that in Ref.~\cite{Moulin:2019ekf} the real part of QNMs has a non-monotonic behavior in terms of $P$. There is no discrepancy in the results; rather, this is all related to the use of different parameters: In our work we fixed $M$ in Fig.~\ref{Rs_QNMS} yielding a variable $m$, while in Ref.~\cite{Moulin:2019ekf} $m$ has been held constant resulting in a variable physical mass $M$. The same can be said for the seemingly contradictory behavior in Fig.~5 of Ref.~\cite{Moulin:2019ekf}, where minus the imaginary part of of QNMs, $-\omega_{\Im }$, appears to be a decreasing function of $\omega_{\Re }$, and in our Fig.~\ref{re2} where $-\omega_{\Im }$ is an increasing function of $\omega_{\Re }$.}
\begin{table}[tbp]
	\begin{tabular}{|l|l|l|l|l|}
		\hline
		\multicolumn{1}{|c|}{ spin 1 } &  \multicolumn{1}{c|}{  $l=1, n=0$ } & \multicolumn{1}{c|}{  $l=2, n=0$ } & \multicolumn{1}{c|}{ $l=2, n=1$ }\\\hline
		$P$ & $\omega \,(WKB)$ & $\omega \,(WKB)$ & $\omega \,(WKB)$   \\ \hline
		0 & 0.2482-0.0926 i & 0.4576-0.0950 i & 0.4365-0.2907 i  \\
		0.1 & 0.3132-0.1280 i & 0.5854-0.1325 i & 0.5501-0.4071 i    \\
		0.2 & 0.3849-0.1696 i & 0.7291-0.1774 i & 0.6744-0.5474 i  \\
		0.3 & 0.4625-0.2163 i & 0.8865-0.2284 i & 0.8082-0.7075 i  \\
		0.4 & 0.5448-0.2665 i & 1.0542-0.2826 i & 0.9510-0.8785 i \ \\\hline
	\end{tabular}
	\caption{The real and imaginary {parts} of QNMs of the electromagnetic field with { $M=1$,  $a_0=0.01$, and} different values of $P$.}\label{tab2}
\end{table}

\begin{figure}
	\includegraphics[width=8 cm]{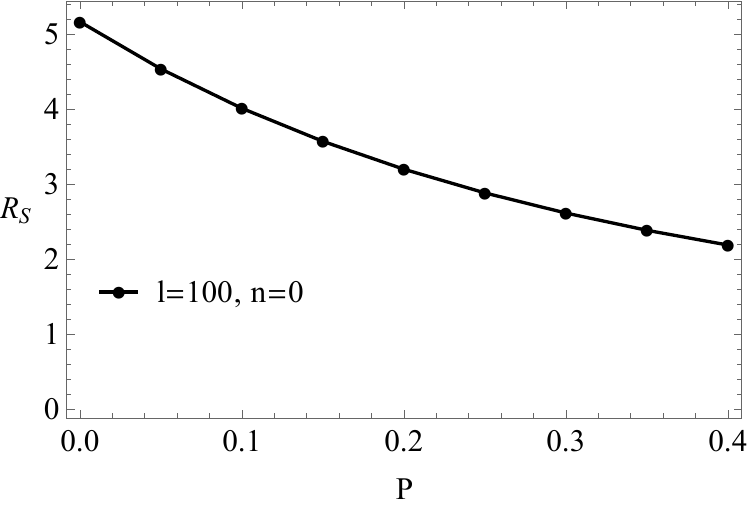}
	\caption{{Shadow radius $R_s$ as a function of $P$ obtained directly from the relation \eqref{k1} by means of the real part of QNMs. In other words, we have estimated  the behavior of shadow radius on $P$ directly from QNMs and without actually solving geodesics equations. One can see that this result is consistent with Fig. (\ref{abcd}) obtained via geodesic equations.} }
	\lb{Rs_QNMS}
\end{figure}

\section{Gravitational deflection of massive particles}

Let us now turn our attention to the gravitational deflection of relativistic massive particles in the spacetime of a LQBH using the approach introduced in Ref.~\cite{Crisnejo:2019ril}. It is worthwhile to mention that the calculation of the deflection angle for different black holes is a very attractive area in the studying of strong gravity features of black holes, see refs. \cite{Kumar:2020hgm, Zhu:2019ura, Bjerrum-Bohr:2014zsa, Younas:2015sva, Azreg-Ainou:2017obt} for examples. To calculate the deflection angle of massive particles, we shall use a correspondence between the motion of a photon in a cold non-magnetized plasma and the motion of a relativistic massive particle in the curved spacetime. Let  $(\mathcal{M},g_{\alpha\beta})$ be a stationary spacetime in the presence of a cold non-magnetized plasma characterized by the refractive index $n$
\begin{equation}\label{refra-index}
	n^2(x,\Omega(x))=1-\frac{\Omega_e^2(x)}{\Omega^2(x)},
\end{equation}
where $\Omega(x)$ is the photon frequency measured by an observer following a timelike Killing vector field. In addition, $\Omega_e(x)$ is known as the plasma frequency given by
\begin{equation}\label{K_e}
	\Omega_e^2(x)=\frac{4\pi e^2}{m_e} N(x)= K_e N(x),
\end{equation}
with $e$ and $m_e$ being the charge {and mass} of the electron, while the quantity $N(x)$ gives the number density of electrons in the plasma media. On the other hand,
 $\Omega(x)$ can be expressed in terms of the gravitational redshift as follows
\begin{equation}
	\Omega(x)=\frac{\Omega_\infty}{\sqrt{-g_{00}}}.
\end{equation}

To {obtain}  the above correspondence we can identify the electron frequency of the plasma $\hbar\Omega_e$ with the rest mass $m$ {and the total energy $E=\hbar \Omega_\infty$ of a photon with the relativistic particle total energy}
\begin{equation}
	E_\infty=\frac{m}{(1-v^2)^{1/2}}.
\end{equation}

We can now {proceed with the calculation of} the deflection angle by applying the geometric method based on the application of the Gauss-Bonnet theorem over the optical geometry. Let $D\subset S$ be a regular domain of the two-dimensional surface $S$ on the optical metric, then the Gauss-Bonnet theorem can be stated as follows
\begin{equation}
	\int\int_D \mathcal{K}dS+\int_{\partial D} k_g dl +\sum_i \epsilon_i = 2\pi\chi(D),
\end{equation}
where $\chi(D)$ is the Euler characteristic number and $\mathcal{K}$ is the Gaussian
curvature, $k_g$ is known as the geodesic curvature of the optical domain $\partial D$, and finally  $\epsilon_i$ gives the corresponding exterior angle in the i-th vertex.  The Gauss-Bonnet theorem
{(GBT)} then can be reformulated in terms of the deflection angle $\hat{\alpha}$  (see, \cite{Crisnejo:2019ril})
\begin{equation}
	\int_0^{\pi+\hat{\alpha}}\bigg[\kappa_g \frac{d\sigma}{d\phi}\bigg]\bigg|_{C_R}d\phi+\int\int_{D_R}\mathcal{K}dS + \int_{\gamma_p} k_g dl =\pi,
\end{equation}
where the limit $R\to\infty$ was applied.
It is well known that for  asymptotically flat spacetimes the following condition is satisfied
\begin{equation}
	\Big[k_g\frac{d\sigma}{d\phi}\Big]_{C_R}\to 1,
\end{equation}
in the limit when the radius of $C_R$ tends to infinity. The GBT then can be simplified as
\begin{equation}\label{alpha-general}
	\hat{\alpha}=-\int\int_{D_R}\mathcal{K}dS - \int_{\gamma_p} k_g dl.
\end{equation}
From the last equation we see that in order to evaluate the deflection angle we need to compute $\mathcal{K}$ and $k_g$, respectively.
In particular, if we simplify the problem further by considering a light {ray} moving in the equatorial plane with $\theta=\pi/2$, then the geodesics  curvature reads,
\begin{equation}\label{eq:kgasada}
	k_g =-\frac{1}{\sqrt{\hat{g} \hat{g}^{\theta\theta} }} \partial_r\hat{\beta}_\phi,
\end{equation}
where $\hat{g}$ {denotes}  the determinant of $\hat{g}_{ab}$. Next, the LQBH metric in the linear order of $a$ can be written as,
\begin{equation}\label{eq:metricoriginal}
	\begin{split}
		ds^2&=f(r)\,dt^2-\frac{dr^2}{g(r)}+\frac{4 a m}{r} dtd\phi -h(r) d\phi^2.
	\end{split}
\end{equation}
By making use of the above correspondence {of the metrics}, one can easily find the Finsler-Randers metric determined by (see, \cite{Crisnejo:2019ril})
\begin{equation}
	 \mathcal{F}(r,\phi,\dot{r},\dot{\phi})=\sqrt{n^2(r)\left[\frac{\dot{r}^2}{f(r)g(r)}+\frac{h(r)}{f(r)}\dot{\phi}^2\right] }- \frac{2 m\, a}{r \,f(r)} \dot{\phi},
\end{equation}
where in the last two equations we have introduced the parameter $m$,  which is related to the ADM mass $M$ by the following relation
\begin{equation}
	m=\frac{M}{(1+P)^2}.
\end{equation}
Furthermore,  the refractive index is given by
\begin{eqnarray}
	n^2(r)&=& 1-(1-v^2)f(r).
\end{eqnarray}
With {these} results in mind, the deflection angle {can be} expressed as follows
\begin{equation}\label{alpha-pm}
	\hat{\alpha}_{\text{mp}}=-\iint_{D_r}\mathcal{K}dS-\int_{R}^S k_g dl.
\end{equation}
{Note that $l$ is {an} affine parameter \cite{Crisnejo:2019ril}, and {$S$ and $R$ represent the source and} receiver, respectively.} Calculating the Gaussian optical curvature {to} leading order terms we find
\begin{equation}
	\mathcal{K}\simeq -\frac{m}{r^3 v^4 }\bigg[ (P+1)^2v^2 +(P-1)^2\bigg]-\frac{8 (1+v^2) a_0^2}{r^6 v^4}.
\end{equation}
On the other hand, for the geodesic curvature contribution it follows
\begin{equation}
	k_g \simeq -\frac{2 a m}{v^2 r^3}.
\end{equation}

If we evaluate this expression using the light ray equation we obtain
\begin{equation}
	\Big[k_gdl\Big]_{r_{\gamma}} \simeq -\frac{2 a m}{v b^2}\sin\phi d\phi,
\end{equation}
{where $b$ denotes the impact parameter for motion of massive particles.} The deflection angle then is,
\begin{equation}\label{alpha-pm}
	\hat{\alpha}_{\text{mp}}=-\int_{0}^{\pi}\int_{b/\sin(\phi)}^{\infty}\mathcal{K} \sqrt{\det \hat{g}} dr d\phi-\int_{0}^{\pi}s\frac{2 a m \sin\phi \, d\phi}{v  b^2},
\end{equation}
where $s= \pm 1$ correspond to the prograde and retrograde orbits respectively. {For prograde orbits the spin of the black hole is in opposite direction of the particle motion, while for retrograde orbit the black hole's spin and the particle's motion are aligned in same direction}. Using our expression \eqref{alpha-pm} we obtain the deflection angle {to the leading order terms,}
\begin{eqnarray}\notag
	\hat{\alpha}_{\text{mp}} &=& \frac{2M}{b}\left[1+\frac{1}{v^2}\frac{(P-1)^2}{(P+1)^2}\right]+\frac{3 \pi a_0^2}{4 b^4}\left(1+\frac{1}{v^2}\right)\\
	&-& \frac{4sa M }{b^2 v}
\end{eqnarray}

Finally, { we  consider a Taylor} expansion around $P$ to obtain the deflection angle
\begin{eqnarray}\notag
	\hat{\alpha}_{\text{mp}} &=& \frac{2M}{b}\left(1+\frac{1}{v^2}\right)-\frac{8MP}{v^2 b}+\frac{3 \pi a_0^2}{4 b^4}\left(1+\frac{1}{v^2}\right)\\
	&-& \frac{4sa M }{b^2 v}
\end{eqnarray}
{Setting} $v=1$,  the deflection angle of light reads
\begin{eqnarray}\notag
	\hat{\alpha}_{\text{light}} &=& \frac{4M}{b}-\frac{8MP}{b}+\frac{3 \pi a_0^2}{2 b^4}- \frac{4sa M }{b^2}
\end{eqnarray}
{which is the same as} the result reported in Ref.~\cite{Sahu:2015dea} for the non-rotating case by setting $a_0=a=0$. From Fig.~\ref{Def} {we find that the deflection angle of massive particles decreases as $P$ increases.}
From the last two equations we see that {taking}  $a_0\to 0$ and $P \to 0$ we obtain the deflection angle of massive particles and light in the Kerr spacetime \cite{Crisnejo:2019ril}. {It is worth noting that in Ref.~\cite{Bjerrum-Bohr:2014zsa} a leading order term due to the quantum gravity effects of the order $G^2\hbar M/b^3$ has been obtained, whereas in the present paper we obtained a leading order term of the order of $8MP/b$. This shows that the quantum effects in the loop quantum gravity are much stronger. Such significant effects are also apparent in the shadow radius as well as in the QNMs. These results can be potentially used in the future to constrain the parameter $P$.}
\begin{figure}
	\includegraphics[width=8.2cm]{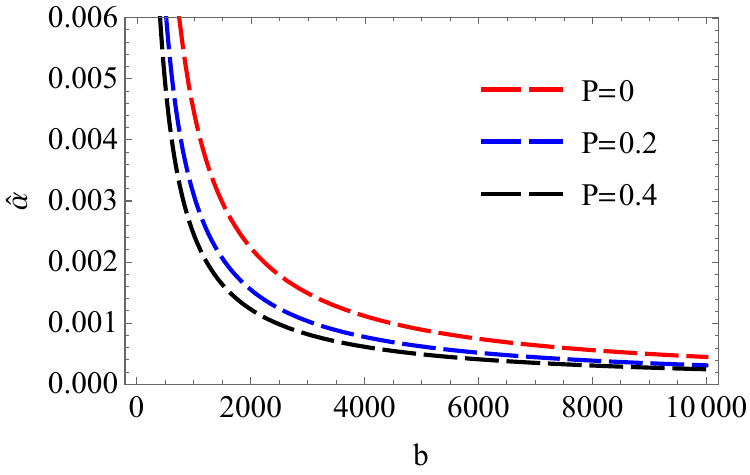}
	\caption{Deflection angle of massive particles for different values of the polymeric function $P$. We have chosen {$M=1$, $a_0=0.01$, $v=0.9$ and  $a=0.8$}. The deflection angle in a rotating LQBH  spacetime is smaller compared to the Kerr Black hole (red curve).}
	\lb{Def}
\end{figure}

\section{Conclusions}

In this paper we have constructed an effective rotating LQBH solution, starting from the spherical symmetric LQBH by applying the Newman-Janis algorithm modified by Azreg-A\"{i}nou's non-complexification procedure~\cite{A}. We have then studied the effects of LQG on its shadow. We have also investigated the physical properties of its horizon, ergosurface, and regularity at $r \simeq 0$, and shown that the black hole solution may have two horizons, one merging horizon, or even no {horizons at all, depending on the polymeric function $P$ and the rotation parameter $a$}. These three cases correspond to the regular rotating LQBH, extreme rotating LQBH, and a regular non-black-hole solutions without {horizons}, respectively.

We have also discussed the effects of {LQG} on both the shape and size of the shadow, {and} shown that, in addition to {the angular} momentum $a$, the polymeric function $P$ causes deformations for both the size and shape of the black hole shadow. For a given value of  {the angular momentum $a$ and the} inclination angle $\theta_0$, the presence of the polymeric function $P$ shrinks the shadow and enhanced its deformation with respect to the shadow {of}  the Kerr spacetime. In addition, we have also {discussed} the effects of the polymeric function $P$ on the deviation from circularity of the shadow, which shows that the deviation from circularity increases with  increasing  $P$ for fixed values of $a$ and $\theta_0$.

In addition,  we have studied the connection between the real part of QNMs in the eikonal limit and {the} shadow radius. First, using the WKB approximation to the sixth order we have shown that the
{QNM} frequencies in the spacetime of {the} LQBH deviate from those of {the} Schwarzschild black hole, as a result, the field perturbations decay more rapidly compared to {the} Schwarzschild black hole due to the induced quantum effects.  Importantly, it is shown that $\omega_{\Re }$ increases with increasing $P$, implying a decrease of the shadow radius $R_S$ due to the inverse relation
\begin{equation}
	R_S(P)=\lim_{l>>1} \frac{l}{\omega_{\Re }(P)}.
\end{equation}
Although this correspondence is precise only in the large limit of $l>>1$, it can still be very useful even in {the} case of small multipoles $l$ to analyze the dependence of $R_S(P)$ on $P$. {This relation has the advantage of predicting the dependence of the shadow radius on $P$ directly from the real part of the QNMs, as is shown in Fig.~\ref{Rs_QNMS}}. This result is shown to be consistent with the one obtained via the geodesic approach as can be seen from Figs.~\ref{a1} and~\ref{bcd}.

Finally, we have also calculated the deflection angle of relativistic massive particles in the spacetime of {the} LQBH. Our calculations show that {the}
deflection angle decreases compared to the Schwarzschild black hole (see from Fig.~\ref{Def}). This result could be {potentially interesting} to distinguish different spacetime solutions.

\section*{Acknowledgements}

C.L. and T.Z. are supported in part by National Natural Science Foundation of China with the Grants No.11675143, the Zhejiang Provincial Natural Science Foundation of China under Grant No. LY20A050002, and the Fundamental Research Funds for the Provincial Universities of Zhejiang in China under Grants No. RF-A2019015.  A.W. is supported by National Natural Science Foundation of China with the Grants Nos. 11675145 and 11975203.

\appendix
	
	\section{The expressions of $r_{h1}$, $r_{h2}$, $r_{e1}$, and $r_{e2}$}
	\begin{widetext}
		\bqn
		r_{h1}&=&\frac{1}{2} \Bigg[\frac{-4 \left(r_++r_-\right) \left(a^2+r_+ r_-\right)-16 a^2 r_*+\left(r_++r_-\right)^3}{4 \sqrt{\frac{1}{3} \left(a^2+r_+ r_-\right)-a^2+\frac{1}{4} \left(r_++r_-\right)^2-r_+r_-+\frac{U}{3 \sqrt[3]{2}}+V}} \nb\\
		&&~~~~~~ -\frac{1}{3} \left(a^2+r_+ r_-\right)-a^2 -V+\frac{1}{2} \left(r_++r_-\right)^2-r_+ r_--\frac{U}{3 \sqrt[3]{2}}\Bigg]^{\frac{1}{2}}, \lb{A1}\\
		r_{h2}&=&\frac{1}{2} \left[\frac{1}{3} \left(a^2+r_+ r_-\right)-a^2+\frac{1}{4} \left(r_++r_-\right)^2-r_+ r_-+\frac{U}{3 \sqrt[3]{2}}+V\right]^{\frac{1}{2}},  \lb{A2}
		\eqn
		where
		\bqn
		U&=&\left[108 a^4 r_*^2-72 a^2 r_*^2 \left(a^2+r_+ r_-\right)+18 a^2 \left(r_++r_-\right) r_* \left(a^2+r_+ r_-\right) \right.\nb\\
		&&\left.+2 \left(a^2+r_+ r_-\right)^3 +27 a^2 \left(r_++r_-\right)^2 r_*^2+W \right]^{\frac{1}{3}},\\
		V&=& \frac{\sqrt[3]{2}}{3 U}\left(12 a^2 r_*^2+6 a^2 \left(r_++r_-\right) r_*+\left(a^2+r_+ r_-\right)^2\right),\\
		W&=&\Big\{\big[108 a^4 r_*^2-72 a^2 r_*^2 \left(a^2+r_+ r_-\right)+18 a^2 \left(r_++r_-\right) r_* \left(a^2+r_+ r_-\right) \nb\\
		&&+2 \left(a^2+r_+ r_-\right)^3+27 a^2 \left(r_++r_-\right)^2 r_*^2\big]^2 \nb\\
		&&-4 \left(12 a^2 r_*^2+6 a^2 \left(r_++r_-\right) r_*+\left(a^2+r_+ r_-\right)^2\right)^3\Big\}^{\frac{1}{2}}.
		\eqn
		
		The expressions of $r_{e1}$ and $r_{e2}$ are given by
		\bqn
		r_{e1}=&&\frac{1}{2} \left[-a^2 \cos ^2\theta+\frac{1}{3} \left(a^2 \cos ^2\theta+r_+ r_-\right)+\frac{1}{4} \left(r_++r_-\right)^2-r_+ r_-+\frac{\mathcal{W}}{3 \sqrt[3]{\mathcal{U}+V}}+\frac{\sqrt[3]{\mathcal{U}+V}}{3 \sqrt[3]{2}}\right]^{\frac{1}{2}},  \lb{A6}\\
		r_{e2}=&&\frac{1}{2} \left[\frac{1}{8 r_{\text{e1}}}\left(\left(r_++r_-\right)^3-4 \left(r_++r_-\right) \left(a^2 \cos ^2\theta+r_- r_+\right)-16 a^2 r_* \cos ^2\theta \right)-a^2 \cos ^2\theta\right. \nb\\
		&&\left.-\frac{1}{3} \left(a^2 \cos ^2\theta+r_+ r_-\right)+\frac{1}{2} \left(r_++r_-\right)^2-r_+r_--\frac{\mathcal{W}}{3 \sqrt[3]{\mathcal{U}+V}}-\frac{\sqrt[3]{\mathcal{U}+V}}{3 \sqrt[3]{2}}\right]^{\frac{1}{2}}, \lb{A7}
		\eqn
		where
		\bqn
		\mathcal{U}=&&\Big\{\Big(108 a^4 r_*^2 \cos ^4\theta+27 a^2 \left(r_{+} + r_{-}\right)^2 r_*^2 \cos ^2\theta - 72 a^2 r_*^2 \cos ^2\theta \left(a^2 \cos ^2\theta +r_+ r_-\right) \nb\\
		&&+18 a^2 r_* \cos ^2\theta\left(r_++r_-\right) \left(a^2 \cos ^2\theta+r_+r_-\right)+2 \left(a^2 \cos ^2\theta +r_+ r_-\right)^3\Big)^2 \nb\\
		&& -4 \left[12 a^2 r_*^2 \cos ^2\theta+6 a^2 \left(r_++r_-\right) r_* \cos ^2\theta+\left(a^2 \cos ^2\theta+r_- r_+\right)^2\right]^3\Big\}^{\frac{1}{2}},   \nb\\  \nb\\
		\mathcal{V}=&&108 a^4 r_*^2 \cos ^4\theta +27 a^2 \left(r_++r_-\right)^2 r_*^2 \cos ^2\theta-72 a^2 r_*^2 \cos ^2\theta \left(a^2 \cos ^2\theta+r_+ r_-\right)\nb\\
		&&+18 a^2 \left(r_++r_-\right) r_* \cos ^2\theta \left(a^2 \cos ^2\theta+r_+ r_-\right)+2 \left(a^2 \cos ^2\theta +r_+ r_-\right)^3,\nb\\  \nb\\
		\mathcal{W}=&&\sqrt[3]{2} \left(12 a^2 r_*^2 \cos ^2\theta +6 a^2 \left(r_++r_-\right) r_* \cos ^2\theta +\left(a^2 \cos ^2\theta +r_+ r_-\right)^2\right). \nb
		\eqn
	\end{widetext}


\begin{thebibliography}{199}
		
		\bibitem{m87}
		K.~Akiyama {\it et al.} [Event Horizon Telescope Collaboration],
		``First M87 Event Horizon Telescope Results. I. The Shadow of the Supermassive Black Hole,''
		\href{\doibase 10.3847/2041-8213/ab0ec7}{Astrophys. J. {\bf 875}, L1 (2019)}.
		
		\bibitem{Akiyama:2019brx}
		K.~Akiyama {\it et al.} [Event Horizon Telescope Collaboration],
		``First M87 Event Horizon Telescope Results. II. Array and Instrumentation,''
		\href{\doibase 10.3847/2041-8213/ab0c96}{Astrophys.\ J.\  {\bf 875}, no. 1, L2 (2019)}
		[arXiv:1906.11239 [astro-ph.IM]].
		
		\bibitem{Akiyama:2019sww}
		K.~Akiyama {\it et al.} [Event Horizon Telescope Collaboration],
		``First M87 Event Horizon Telescope Results. III. Data Processing and Calibration,''
		\href{\doibase 10.3847/2041-8213/ab0c57}{Astrophys.\ J.\  {\bf 875}, no. 1, L3 (2019)}
		[arXiv:1906.11240 [astro-ph.GA]].
		
		\bibitem{Akiyama:2019bqs}
		K.~Akiyama {\it et al.} [Event Horizon Telescope Collaboration],
		``First M87 Event Horizon Telescope Results. IV. Imaging the Central Supermassive Black Hole,''
		\href{\doibase 10.3847/2041-8213/ab0e85}{Astrophys.\ J.\  {\bf 875}, no. 1, L4 (2019)}
		[arXiv:1906.11241 [astro-ph.GA]].
		
		\bibitem{Akiyama:2019fyp}
		K.~Akiyama {\it et al.} [Event Horizon Telescope Collaboration],
		``First M87 Event Horizon Telescope Results. V. Physical Origin of the Asymmetric Ring,''
		\href{\doibase 10.3847/2041-8213/ab0f43}{Astrophys.\ J.\  {\bf 875}, no. 1, L5 (2019)}
		[arXiv:1906.11242 [astro-ph.GA]].
		
		\bibitem{Akiyama:2019eap}
		K.~Akiyama {\it et al.} [Event Horizon Telescope Collaboration],
		``First M87 Event Horizon Telescope Results. VI. The Shadow and Mass of the Central Black Hole,''
		\href{\doibase 10.3847/2041-8213/ab1141}{Astrophys. J. {\bf 875}, L6 (2019)}.
		
		\bibitem{ngVLA}
		A. M. Hughes, A. Beasley, and C. Carilli,
		``Next Generation Very Large Array: Centimeter Radio Astronomy in the 2020s," %
		IAU General Assembly {\bf 22}, 2255106 (2015).
		
		\bibitem{TMT}
		G. H. Sanders, %
		``The Thirty Meter Telescope (TMT): An International Observatory," %
		\href{\doibase 10.1007/s12036-013-9169-5}{Journal of Astrophysics and Astronomy {\bf 34}, 81 (2013)}.
		
		\bibitem{BlackHoleCam}
		C.~Goddi {\it et al.},
		``BlackHoleCam: Fundamental physics of the galactic center,''
		\href{\doibase 10.1142/S0218271817300014}{Int. J. Mod. Phys. D {\bf 26}, 1730001 (2016)}.
		
		\bibitem{Claudel:2000yi}
		C.~M.~Claudel, K.~S.~Virbhadra and G.~F.~R.~Ellis,
		``The Geometry of photon surfaces,''
		\href{\doibase 10.1063/1.1308507}{J.\ Math.\ Phys.\  {\bf 42}, 818 (2001)}.
		
		\bibitem{wang_shadow_2019}
		M. Wang, S. Chen, J. Wang, and J. Jing, %
		``Shadow of a Schwarzschild black hole surrounded by a Bach-Weyl ring," %
		\href{http://arxiv.org/abs/1904.12423}{arXiv:1904.12423 [gr-qc]}.
		
		
		
		\bibitem{gb}
		P.V. Cunha, C.A.R. Herdeiro, B. Kleihaus, J. Kunz, and E. Radu, %
		``Shadows of Einstein-dilaton-Gauss-Bonnet black holes,"
		\href{\doibase 10.1016/j.physletb.2017.03.020}{Phys. Lett. {\bf B 768}, 373 (2017)}.
		
		\bibitem{kz}
		M. Wang, S. Chen, and J. Jing,
		``Shadow casted by a Konoplya-Zhidenko rotating non-Kerr black hole,"
		\href{\doibase 10.1088/1475-7516/2017/10/051}{\JCAP ~10, 051 (2017)}.
		
		
		\bibitem{Younsi:2016azx}
		Z.~Younsi, A.~Zhidenko, L.~Rezzolla, R.~Konoplya and Y.~Mizuno,
		``New method for shadow calculations: Application to parametrized axisymmetric black holes,''
		\href{\doibase 10.1103/PhysRevD.94.084025}{Phys. Rev. D {\bf 94}, 084025 (2016)}.
		
		\bibitem{cs}
		M. Amir, B.P. Singh, and S.G. Ghosh,
		``Shadows of rotating five-dimensional charged EMCS black holes, "
		\href{\doibase 10.1140/epjc/s10052-018-5872-3}{Eur. Phys. J. C {\bf 78}, 399 (2018)}.
		
		\bibitem{wei_observing_2013}
		S.-W. Wei and Y.-X. Liu, %
		``Observing the shadow of Einstein-Maxwell-Dilaton-Axion black hole,"
		\href{\doibase 10.1088/1475-7516/2013/11/063}{J. Cosmol. Astropart. Phys. 2013, 063 (2013)}.
		
		\bibitem{Amir:2018pcu}
  M.~Amir, K.~Jusufi, A.~Banerjee and S.~Hansraj, ``Shadow images of Kerr-like wormholes,''
\href{\doibase 10.1088/1361-6382/ab42be}{Class.\ Quant.\ Grav.\  {\bf 36} (2019) no.21,  215007}.
		
		\bibitem{pfdm}
		S. Haroon, M. Jamil, K. Jusufi, K. Lin, and R.B. Mann, %
		``Shadow and deflection angle of rotating black holes in perfect fluid dark matter with a cosmological constant,"
		\href{\doibase 10.1103/PhysRevD.99.044015}{Phys. Rev. D {\bf 99}, 044015 (2019)}.
		
		\bibitem{kds}
		Z. Stuchlik, D. Charbulak, and J. Schee, %
		``Light escape cones in local reference frames of Kerr Ã¢Â€Â“de Sitter  black hole spacetimes and related black hole shadows,"
		\href{\doibase 10.1140/epjc/s10052-018-5578-6}{Eur. Phys. J. C 78, 180 (2018)}.
		
		\bibitem{mog}
		J.W. Moffat, %
		``Black holes in modified gravity (MOG), "%
		\href{\doibase 10.1140/epjc/s10052-015-3405-x}{Eur. Phys. J. C 75, 175 (2015)}.
		
		\bibitem{wang_shadows_2019}
		H.-M. Wang, Y.-M. Xu, and S.-W. Wei, %
		``Shadows of Kerr-like black holes in a modified gravity theory, " %
		\href{\doibase 10.1088/1475-7516/2019/03/046}{J. Cosmol. Astropart. Phys. 2019, 046 (2019)}.
		
		
		\bibitem{rr} A. Abdujabbarov, M. Amir, B. Ahmedov, S.G. Ghosh, %
		``Shadow of rotating regular black holes, "%
		\href{\doibase 10.1103/PhysRevD.93.104004}{Phys. Rev. D {\bf 93}, 104004 (2016)}.
		
		\bibitem{Dastan:2016bfy}
		S.~Dastan, R.~Saffari and S.~Soroushfar,
		``Shadow of a Kerr-Sen dilaton-axion Black Hole,''
		\href{https://arxiv.org/abs/1610.09477v1}{arXiv:1610.09477 [gr-qc]}.
		
		\bibitem{ks}
		S.~Dastan, R.~Saffari and S.~Soroushfar,
		``Shadow of a Charged Rotating Black Hole in $f(R)$ Gravity,''
		\href{https://arxiv.org/abs/1606.06994v1}{arXiv:1606.06994 [gr-qc]}.
		
		\bibitem{nc}
		M. Sharif, S. Iftikhar, %
		``Shadow of a charged rotating non-commutative black hole, "
		\href{\doibase 10.1140/epjc/s10052-016-4472-3}{Eur. Phys. J. C {\bf 76}, 630 (2016)}.
		
		\bibitem{Bambi:2010hf}
		C.~Bambi and N.~Yoshida,
		``Shape and position of the shadow in the $\delta = 2$ Tomimatsu-Sato space-time,'' %
		\href{\doibase 10.1088/0264-9381/27/20/205006}{Class. Quant. Grav. {\bf 27}, 205006 (2010)}.
		
		\bibitem{Jusufi:2019nrn}
		K.~Jusufi, M.~Jamil, P.~Salucci, T.~Zhu, and S.~Haroon,
		``Black Hole Surrounded by Dark Matter Halo in the M87 Galactic Center and its Identification with Shadow Images,''
		\href{\doibase 10.1103/PhysRevD.100.044012}{Phys.\ Rev.\ D {\bf 100}, 044012 (2019)}.
		
		\bibitem{hou_black_2018}
		X. Hou, Z. Xu, M. Zhou, and J. Wang, %
		``Black hole shadow of SgrA$^\ast$ in dark matter halo, "%
		\href{\doibase 10.1088/1475-7516/2018/07/015}{J. Cosmol. Astropart. Phys. 2018, 015 (2018)}.
		
		\bibitem{konoplya_shadow_2019}
		R. A. Konoplya, %
		Shadow of a black hole surrounded by dark matter, %
		\href{http://arxiv.org/abs/1905.00064}{arXiv: 1905.00064 [gr-qc]}.
		
		
		\bibitem{Haroon:2019new}
        S.~Haroon, K.~Jusufi and M.~Jamil,
		``Shadow Images of a Rotating Dyonic Black Hole with a Global Monopole Surrounded by Perfect Fluid,"
		\href{\doibase 10.3390/universe6020023}{Universe {\bf 6} (2020) no.2,  23}.
		
		
		\bibitem{ns}
		R. Shaikh, P. Kocherlakota, R. Narayan, and P.S. Joshi, %
		``Shadows of spherically symmetric black holes and naked singularities,"
		\href{\doibase 10.1093/mnras/sty2624}{MNRAS 482, 52 (2019)}.
		
		\bibitem{Bambi:2008jg}
		C.~Bambi and K.~Freese,
		``Apparent shape of super-spinning black holes,'' %
		\href{\doibase 10.1103/PhysRevD.79.043002}{Phys. Rev. D {\bf 79}, 043002 (2009)}.
		
		\bibitem{wei_intrinsic_2019}
		S.-W. Wei, Y.-X. Liu, and R. B. Mann, %
		Intrinsic curvature and topology of shadows in Kerr spacetime, %
		\href{\doibase 10.1103/PhysRevD.99.041303}{Phys. Rev. D {\bf 99}, 041303 (2019)}.
		
		\bibitem{wei_curvature_2019}
		S.-W. Wei, Y.-C. Zou, Y.-X. Liu, and R. B. Mann, %
		Curvature radius and Kerr black hole shadow, %
		\href{http://arxiv.org/abs/1904.07710}{arXiv:1904.07710 [gr-qc]}.
		
		\bibitem{Li:2020zxi}
		Z.~Li and T.~Zhou,
		``Kerr black hole surrounded by a cloud of strings and its weak gravitational lensing in Rastall gravity,''
		\href{https://arxiv.org/abs/2001.01642}{arXiv:2001.01642 [gr-qc]}.
		
		\bibitem{Li:2019lsm}
		C.~Li, S.~F.~Yan, L.~Xue, X.~Ren, Y.~F.~Cai, D.~A.~Easson, Y.~F.~Yuan and H.~Zhao,
		``Testing the equivalence principle via the shadow of black holes,''
		\href{https://arxiv.org/abs/1912.12629}{arXiv:1912.12629 [astro-ph.CO]}.
		
		\bibitem{Konoplya:2019xmn}
		R.~A.~Konoplya,
		``Quantum corrected black holes: quasinormal modes, scattering, shadows,''
		\href{https://arxiv.org/abs/1912.10582}{arXiv:1912.10582 [gr-qc]}.
		
		\bibitem{Allahyari:2019jqz}
		A.~Allahyari, M.~Khodadi, S.~Vagnozzi and D.~F.~Mota,
		``Magnetically charged black holes from non-linear electrodynamics and the Event Horizon Telescope,''
		\href{https://arxiv.org/abs/1912.08231}{arXiv:1912.08231 [gr-qc]}.
		
		\bibitem{Dokuchaev:2019jqq}
		V.~I.~Dokuchaev and N.~O.~Nazarova,
		``Silhouettes of invisible black holes,''
		\href{https://arxiv.org/abs/1911.07695}{arXiv:1911.07695 [gr-qc]}.
		
		\bibitem{Jusufi:2019caq}
		K. Jusufi, M. Jamil, H. Chakrabarty, Q. Wu, C. Bambi, and A. Wang,
        ``Rotating regular black holes in conformal massive gravity'', \href{https://arxiv.org/abs/1911.07520}{arXiv:1911.07520 [gr-qc]}, accepted for publication in Phys. Rev. D.
    
        \bibitem{Ding:2019mal}
        C.~Ding, C.~Liu, R.~Casana and A.~Cavalcante,
        ``Exact Kerr-Like Solution and Its Shadow in a Gravity Model with Spontaneous Lorentz Symmetry Breaking,''
        \href{https://arxiv.org/abs/1910.02674}{arXiv:1910.02674 [gr-qc]}.
        
        \bibitem{Konoplya:2019fpy}
        R.~A.~Konoplya, T.~Pappas and A.~Zhidenko,
        ``Einstein--scalar--Gauss--Bonnet black holes: Analytical approximation for the metric and applications to calculations of shadows,''
        \href{https://arxiv.org/abs/1907.10112}{arXiv:1907.10112 [gr-qc]}.
        
        \bibitem{Konoplya:2019goy}
        R.~A.~Konoplya and A.~Zhidenko,
        ``Analytical representation for metrics of scalarized Einstein-Maxwell black holes and their shadows,''
        \href{\doibase 10.1103/PhysRevD.100.044015}{Phys.\ Rev.\ D {\bf 100}, 044015 (2019)}
        [arXiv:1907.05551 [gr-qc]].
        
        \bibitem{Vagnozzi:2020quf}
        S.~Vagnozzi, C.~Bambi and L.~Visinelli,
        ``Concerns regarding the use of black hole shadows as standard rulers,''
        \href{https://arxiv.org/abs/2001.02986}{arXiv:2001.02986 [gr-qc]}.
        
        \bibitem{Yu:2020bxd}
        S.~Yu and C.~Gao,
        ``An exact black hole spacetime with scalar field and its shadow together with quasinormal modes,''
        \href{https://arxiv.org/abs/2001.01137}{arXiv:2001.01137 [gr-qc]}.
        
        \bibitem{Kumar:2020hgm}
        R.~Kumar, S.~G.~Ghosh and A.~Wang,
        ``Light deflection and shadow cast by rotating Kalb-Ramond black holes,''
        \href{https://arxiv.org/abs/2001.00460}{arXiv:2001.00460 [gr-qc]}.
        
        \bibitem{Tian:2019yhn}
        S.~X.~Tian and Z.~H.~Zhu,
        ``Testing the Schwarzschild metric in a strong field region with the Event Horizon Telescope,''
        \href{\doibase 10.1103/PhysRevD.100.064011}{Phys.\ Rev.\ D {\bf 100}, 064011 (2019)}
        [arXiv:1908.11794 [gr-qc]].
        
        \bibitem{Azreg:2015}
        M.~Azreg-A\"{i}nou,
        ``Confined-exotic-matter wormholes with no gluing effects --- Imaging supermassive wormholes and black
        holes,''
        \href{\doibase 10.1088/1475-7516/2015/07/037}{JCAP07, 037 (2015)}.
        
        \bibitem{Zhu:2019ura}
        T.~Zhu, Q.~Wu, M.~Jamil and K.~Jusufi,
        ``Shadows and deflection angle of charged and slowly rotating black holes in Einstein-\AE{}ther theory,''
        \href{\doibase 10.1103/PhysRevD.100.044055}{Phys.\ Rev.\ D {\bf 100}, 044055 (2019)}
        [arXiv:1906.05673 [gr-qc]].
        
        \bibitem{Bambi:2019tjh}
        C.~Bambi, K.~Freese, S.~Vagnozzi and L.~Visinelli,
        ``Testing the rotational nature of the supermassive object M87* from the circularity and size of its first image,''
        \href{\doibase 10.1103/PhysRevD.100.044057}{Phys.\ Rev.\ D {\bf 100}, no. 4, 044057 (2019)}
        [arXiv:1904.12983 [gr-qc]].
        
        \bibitem{Vagnozzi:2019apd}
        S.~Vagnozzi and L.~Visinelli,
        ``Hunting for extra dimensions in the shadow of M87*,''
        \href{\doibase 10.1103/PhysRevD.100.024020}{Phys.\ Rev.\ D {\bf 100}, no. 2, 024020 (2019)}
        [arXiv:1905.12421 [gr-qc]].
        
        \bibitem{Banerjee:2019nnj}
        I.~Banerjee, S.~Chakraborty and S.~SenGupta, ``Silhouette of M87*: A New Window to Peek into the World of Hidden Dimensions,''
        \url{arXiv:1909.09385 [gr-qc]}.
        
     \bibitem{Held:2019xde} 
A.~Held, R.~Gold and A.~Eichhorn,
  ``Asymptotic safety casts its shadow,''
  \href{\doibase:10.1088/1475-7516/2019/06/029}{JCAP {\bf 1906}, 029 (2019)}.
 
        
        \bibitem{LQG_BH}
        L. Modesto, ``Semiclassical Loop Quantum Black Hole,"
        \href{\doibase 10.1007/s10773-010-0346-x}{Int. J. Theor. Phys. 49, 1649 (2010)}.

       \bibitem{erratum}
       Cheng Liu, Tao Zhu, Qiang Wu, Kimet Jusufi, Mubasher Jamil, Mustapha Azreg-Aïnou, and Anzhong Wang,
``Erratum: Shadow and quasinormal modes of a rotating loop quantum black hole, "
\href{https://doi.org/10.1103/PhysRevD.103.089902}{Phys. Rev. D 103, 089902}
        
        \bibitem{AOS18a} A. Ashtekar,  J. Olmedo, and P. Singh, Quantum Transfiguration of Kruskal Black Holes, \href{\doibase  10.1103/PhysRevLett.121.241301}
        {Phys. Rev. Lett. {\bf 121}, 241301 (2018)}.
        
        \bibitem{AOS18b}  A. Ashtekar,  J. Olmedo, and P. Singh, Quantum extension of the Kruskal spacetime, \href{\doibase  10.1103/PhysRevD.98.126003}
        {Phys. Rev. D{\bf 98}, 126003 (2018)}.
        
        \bibitem{BBy18} M. Bojowald, S. Brahma, and D.-H. Yeom, Effective line elements and black-hole models in canonical loop quantum gravity, \href{\doibase  10.1103/PhysRevD.98.046015}{Phys. Rev. D{\bf 98}, 046015 (2018)}; J.~Ben Achour, F.~Lamy, H.~Liu and K.~Noui,
  ``Polymer Schwarzschild black hole: An effective metric,''  \href{\doibase:10.1209/0295-5075/123/20006}{EPL {\bf 123}, no. 2, 20006 (2018)}.
 
        
        \bibitem{ABP19}  E. Alescia, S. Bahramia, D. Pranzetti, Quantum gravity predictions for black hole interior geometry,
        \href{\doibase  10.1016/j.physletb.2019.134908} {Phys. Lett. B{\bf 797} (2019) 134908}.
        
        \bibitem{ADL20}  M. Assanioussi,  A. Dapor,  and K.  Liegener, Perspectives on the dynamics in a loop quantum gravity effective description
        of black hole interiors, \href{\doibase  10.1103/PhysRevD.101.026002} {Phys. Rev. D{\bf 101}, 026002 (2020)}.
        
        \bibitem{Perez17} A. Perez, Black holes in loop quantum gravity, \href{\doibase  10.1088/1361-6633/aa7e14} {Rep. Prog. Phys. {\bf 80} (2017) 126901}.
        
        \bibitem{BMM18}  A. Barrau, K. Martineau and F. Moulin, A Status Report on the Phenomenology of Black Holes in Loop Quantum Gravity:
        Evaporation, Tunneling to White Holes, Dark Matter and
        Gravitational Waves, \href{\doibase  10.3390/universe4100102} {Universe {\bf 4} (2018) 102}.
        
        \bibitem{Rovelli18} C. Rovelli, Black hole evolution traced out with Loop Quantum Gravity, Phys. {\bf 11} (2018) 127[\href{\doibase arxiv.org/pdf/1901.04732.pdf}{arXiv:1901.04732}].
        
        \bibitem{Ashtekar20} A. Ashtekar, Black Hole evaporation: A Perspective from Loop Quantum Gravity, \href{\doibase arxiv.org/pdf/2001.08833.pdf}{arXiv:2001.08833}.
        
        \bibitem{Alesci:2011wn}
        E.~Alesci and L.~Modesto,
        ``Particle Creation by Loop Black Holes,''
        \href{\doibase 10.1007/s10714-013-1656-0}{Gen.\ Rel.\ Grav.\  {\bf 46}, 1656 (2014)}
        [arXiv:1101.5792 [gr-qc]].
        
        \bibitem{Chen:2011zzi}
        J.~H.~Chen and Y.~J.~Wang,
        ``Complex frequencies of a massless scalar field in loop quantum black hole spacetime,''
        \href{\doibase 10.1088/1674-1056/20/3/030401}{Chin.\ Phys.\ B {\bf 20}, 030401 (2011)}.
        
        \bibitem{Dasgupta:2012nk}
        A.~Dasgupta,
        ``Entropy Production and Semiclassical Gravity,''
        \href{\doibase 10.3842/SIGMA.2013.013}{SIGMA {\bf 9}, 013 (2013)}
        [arXiv:1203.5119 [gr-qc]].
        
        \bibitem{Barrau:2014yka}
        A.~Barrau, C.~Rovelli and F.~Vidotto,
        ``Fast Radio Bursts and White Hole Signals,''
        \href{\doibase 10.1103/PhysRevD.90.127503}{Phys.\ Rev.\ D {\bf 90}, 127503 (2014)}
        [arXiv:1409.4031 [gr-qc]].
        
        
        \bibitem{Hossenfelder:2012tc}
        S.~Hossenfelder, L.~Modesto and I.~Premont-Schwarz,
        ``Emission spectra of self-dual black holes,''
        \href{https://arxiv.org/abs/1202.0412}{arXiv:1202.0412 [gr-qc]}.
        
        \bibitem{Sahu:2015dea}
        S.~Sahu, K.~Lochan and D.~Narasimha,
        ``Gravitational lensing by self-dual black holes in loop quantum gravity,''
        \href{\doibase 10.1103/PhysRevD.91.063001}{Phys.\ Rev.\ D {\bf 91}, 063001 (2015)}
        [arXiv:1502.05619 [gr-qc]].
        
        \bibitem{Cruz:2015bcj}
        M.~B.~Cruz, C.~A.~S.~Silva and F.~A.~Brito,
        ``Gravitational axial perturbations and quasinormal modes of loop quantum black holes,''
        \href{\doibase 10.1140/epjc/s10052-019-6565-2}{Eur.\ Phys.\ J.\ C {\bf 79}, 157 (2019)}.
        
        \bibitem{Caravelli:2010ff}
        F.~Caravelli and L.~Modesto,
        ``Spinning Loop Black Holes,''
        \href{\doibase 10.1088/0264-9381/27/24/245022}{Class.\ Quant.\ Grav.\  {\bf 27}, 245022 (2010)}
        [arXiv:1006.0232 [gr-qc]].
        
        \bibitem{AzregAinou:2011fq}
        M.~Azreg-A\"{i}nou,
        ``Comment on `Spinning loop black holes' [arXiv:1006.0232],''
        \href{\doibase:10.1088/0264-9381/28/14/148001}{Class.\ Quant.\ Grav.\  {\bf 28}, 148001 (2011)}
        [arXiv:1106.0970 [gr-qc]].
        
        \bibitem{NJ1}
        E. T. Newman and A. I. Janis,
        ``Note on the Kerr spinning-particle metric'',
        \href{\doibase 10.1063/1.1704350}{Journal of Mathematical Physics \textbf{6}, 915 (1965)}.
        
        \bibitem{NJ2}
        E. T. Newman, et. al.,
        ``Metric of a rotating charged mass'',
        \href{\doibase 10.1063/1.1704351}{Journal of Mathematical Physics \textbf{6}, 918 (1965)}.
        
        \bibitem{Til}
        C. J. Talbot,
        ``Newman-Penrose approach to twisting degenerate metrics'',
        \href{https://doi.org/10.1007/BF01645269}{Communications in Mathematical Physics \textbf{13}, 45 (1969)}.
        
        \bibitem{drake}
        S. P. Drake, S.P. and P. Szekeres,
        ``Uniqueness of the Newman--Janis algorithm in generating the Kerr--Newman metric'',
        \href{\doibase 10.1023/A:1001920232180}{General relativity and Gravitation, \textbf{32}, 445 (2000)}.
        
        		\bibitem{A}
        M. Azreg-A\"{i}nou,
        ``Generating rotating regular black hole solutions without complexification'',
        \href{\doibase 10.1103/PhysRevD.90.064041}{Physical Review D \textbf{90}, 064041 (2014)}.
        
        
        
        
        \bibitem{A1}
        M. Azreg-A\"{i}nou,
        ``From static to rotating to conformal static solutions: rotating imperfect fluid wormholes with(out) electric or magnetic field'',
        \href{\doibase 10.1140/epjc/s10052-014-2865-8}{Eur. Phys. J. C \textbf{74}, 2865 (2014)}.        
        
    
        \bibitem{s18}
        E. Contreras, J. M. Ramirez-Velasquez, \'{A} Rinc\'{o}n, G. Panotopoulos, P. Bargue\~{n}o,
        ``Black hole shadow of a rotating polytropic black hole by the Newman--Janis algorithm without complexification'', \href{\doibase 10.1140/epjc/s10052-019-7309-z}{Eur. Phys. J. C {\bf 79}, 802 (2019)}.
        
        		\bibitem{s1}
        B. Toshmatov, Z. Stuchl\'{\i}k, and B. Ahmedov,
        ``Rotating black hole solutions with quintessential energy'',
        \href{\doibase 10.1140/epjp/i2017-11373-4}{Eur. Phys. J. Plus \textbf{132}, 98 (2017)}.
        
        \bibitem{s2}
        A. Abdujabbarov, B. Toshmatov, Z. Stuchl\'{\i}k, and B. Ahmedov,
        ``Shadow of the rotating black hole with quintessential energy in the presence of plasma'',
        \href{https://doi.org/10.1142/S0218271817500511}{Int. J. Mod. Phys. D \textbf{26}, 1750051 (2017)}.
        
        \bibitem{s3}
        Z.~Xu and J.~Wang,
        ``Kerr--Newman-AdS black hole in quintessential dark energy'',
        \href{\doibase 10.1103/PhysRevD.95.064015}{Phys. Rev. D \textbf{95}, 064015 (2017)}.
        
        \bibitem{s4}
        B. Toshmatov, Z. Stuchl\'{\i}k, and B. Ahmedov,
        ``Comments on \textquotedblleft Casimir effect in the Kerr spacetime with quintessence\textquotedblright '',
        \href{\doibase 10.1142/S0217732317750013}{Mod. Phys. Lett. A \textbf{32}, 1775001 (2017)}.
        
        \bibitem{s5}
        S. Haroon, M. Jamil, K. Lin, P. Pavlovic, M. Sossich, and A. Wang,
        ``The Effects of Running Gravitational Coupling On Rotating Black Holes'',
        \href{\doibase 10.1140/epjc/s10052-018-5986-7}{Eur. Phys. J. C \textbf{78}, 519 (2018)}.
        
        \bibitem{s6}
        Z. Xu, J. Wang,
        ``Kerr--Newman-AdS Black Hole Surrounded By Scalar Field Matter In Rastall Gravity'',
        \href{https://arxiv.org/abs/1711.04542}{arXiv:1711.04542}.
        
        \bibitem{s7}
        B. Toshmatov, Z. Stuchl\'{\i}k, and B. Ahmedov,
        ``Generic rotating regular black holes in general relativity coupled to nonlinear electrodynamics'',
        \href{\doibase 10.1103/PhysRevD.95.084037}{Phys. Rev. D \textbf{95}, 084037 (2017)}.
        
        \bibitem{s8}
        M. Azreg-A\"{i}nou,
        ``Wormhole solutions sourced by fluids, I: Two-fluid charged sources'',
        \href{https://doi.org/10.1140/epjc/s10052-015-3835-5}{Eur. Phys. J. C \textbf{76}, 3 (2016)}.
        
        \bibitem{s9}
        M. Azreg-A\"{i}nou,
        ``Wormhole solutions sourced by fluids, II: three-fluid two-charged sources'',
        \href{\doibase 10.1140/epjc/s10052-015-3836-4}{Eur. Phys. J. C \textbf{76}, 7 (2016)}.
        
        \bibitem{s10}
        Z.~Xu, X.~Hou, and J.~Wang,
        ``Kerr--anti-de Sitter/de Sitter black hole in perfect fluid dark matter background'',
        \href{\doibase 10.1088/1361-6382/aabcb6}{Class. Quantum Grav. \textbf{35}, 115003 (2018)}.
        
        \bibitem{s11}
        C. A. Benavides-Gallego, A. A. Abdujabbarov, and C. Bambi,
        ``Rotating and non-linear magnetic-charged black hole surrounded by quintessence'',
        \href{https://arxiv.org/abs/1811.01562}{arXiv:1811.01562v1 [gr-qc]}.
        
        \bibitem{s12}
        M.~Azreg-A\"{i}nou, S.~Haroon, M.~Jamil, and M.~Rizwan
        ``Rotating normal and phantom Einstein--Maxwell--dilaton black holes: Geodesic analysis'',
        \href{\doibase 10.1142/S0218271819500639}{Int. J. Mod. Phys. D \textbf{28}, 1950063 (2019)}.
        
        \bibitem{s13}
        Z. Xu, X. Gong, and S-N. Zhang,
        ``Black hole immersed dark matter halo'',
        \href{\doibase 10.1103/PhysRevD.101.024029}{Phys. Rev. D \textbf{101}, 024029 (2020)}.
        
        \bibitem{s14}
        M. Sharif and Q. Ama-Tul-Mughani,
        ``Greybody factor for a rotating Bardeen black hole'',
        \href{\doibase 10.1140/epjp/i2019-12979-0}{Eur. Phys. J. Plus \textbf{134}, 616 (2019)}.
        
        \bibitem{s15}
        N. Bret\'{o}n, C. L\"{a}mmerzahl, and A. Mac\'{\i}as,
        ``Rotating black holes in the Einstein--Euler--Heisenberg theory'',
        \href{\doibase 10.1088/1361-6382/ab5169}{Class. Quantum Grav. \textbf{36}, 235022 (2019)}.
        
        \bibitem{s16}
        C-Y. Chen and P. Chen,
        ``Separability of the Klein-Gordon equation for rotating spacetimes obtained from Newman-Janis algorithm'',
        \href{\doibase 10.1103/PhysRevD.100.104054}{Phys. Rev. D \textbf{100}, 104054 (2019)}.
        
        \bibitem{s17}
        Z. Xu, Y. Liao and J. Wang,
        ``Thermodynamics and phase transition in rotational Kiselev black hole'',
        \href{\doibase 10.1142/S0217751X19501859}{Int. J. Mod. Phys. A \textbf{34}, 1950185 (2019)}.
        
        \bibitem{cardoso} V. Cardoso, A. S. Miranda, E. Berti, H. Witek, and V. T. Zanchin, Geodesic stability, Lyapunov exponents,
        and quasinormal modes, \href{\doibase 10.1103/PhysRevD.79.064016}{Phys. Rev. D {\bf 79}, 064016 (2009)}.
        
              
        \bibitem{Stefanov:2010xz}
        I.~Z.~Stefanov, S.~S.~Yazadjiev and G.~G.~Gyulchev,
        ``Connection between Black-Hole Quasinormal Modes and Lensing in the Strong Deflection Limit,''
        \href{\doibase 10.1103/PhysRevLett.104.251103}{Phys.\ Rev.\ Lett.\  {\bf 104}, 251103 (2010)}.
        
          \bibitem{wei2019}
        S.~W.~Wei and Y.~X.~Liu,
  ``Null Geodesics, Quasinormal Modes, and Thermodynamic Phase Transition for Charged Black Holes in Asymptotically Flat and dS Spacetimes,'' \href{http://arxiv.org/abs/1909.11911}{arXiv:1909.11911 [gr-qc]}.

        
        \bibitem{Jusufi:2019ltj}
        K.~Jusufi, ``Quasinormal Modes of Black holes Surrounded by Dark Matter and Their Connection with Shadow Radius,''
        \href{https://arxiv.org/abs/1912.13320}{arXiv:1912.13320 [gr-qc]}.
        
        \bibitem{Konoplya:2017wot}
        R.~A.~Konoplya and Z.~Stuchlík, ``Are eikonal quasinormal modes linked to the unstable circular null geodesics?,''
        \href{\doibase 10.1016/j.physletb.2017.06.015}{ Phys.\ Lett.\ B {\bf 771} (2017) 597}. 
        
        \bibitem{Boonserm:2013dua}
        P.~Boonserm, T.~Ngampitipan and M.~Visser,``Regge-Wheeler equation, linear stability, and greybody factors for dirty black holes,''
        \href{\doibase 10.1103/PhysRevD.88.041502}{Phys.\ Rev.\ D {\bf 88} (2013) 041502}.
        
        \bibitem{Schutz} B. F. Schutz and C. M. Will, Black hole normal modes - A semianalytic approach, \href{\doibase 10.1086/184453}{Astrophys. J. Lett. 291 (1985) L33}.
        
        \bibitem{Iyer} S. Iyer and C. M. Will, Black-hole normal modes: A WKB approach. I. Foundations and application of a
        higher-order WKB analysis of potential-barrier scattering, \href{\doibase 10.1103/PhysRevD.35.3621}{Phys. Rev. D {\bf 35}, 3621 (1987)}.
        
        \bibitem{KonoplyaWKB} R. A. Konoplya, Quasinormal behavior of the $D$-dimensional Schwarzschild black hole and the higher order
        WKB approach, \href{\doibase 10.1103/PhysRevD.68.024018}{Phys. Rev. D {\bf 68},  024018 (2003)}.
        
        \bibitem{Moulin:2019ekf}
        F.~Moulin, A.~Barrau and K.~Martineau, ``An overview of quasinormal modes in modified and extended gravity,''
        \href{\doibase 10.3390/universe5090202}{Universe {\bf 5} (2019) no.9,  202}
        
        
		
		%
		%
		%
		%
		%
		%
		%
		
		
		
		
		%
		%
		%
		%
		%
		%
		
		%
		%
		%
		%
		%
		
		\bibitem{Crisnejo:2019ril}
		G.~Crisnejo, E.~Gallo and K.~Jusufi, ``Higher order corrections to deflection angle of massive particles and light rays in plasma media for stationary spacetimes using the Gauss-Bonnet theorem,''
		\href{\doibase doi:10.1103/PhysRevD.100.104045}{ Phys.\ Rev.\ D {\bf 100} (2019) no.10,  104045}.
		
\bibitem{Bjerrum-Bohr:2014zsa}
  N.~E.~J.~Bjerrum-Bohr, J.~F.~Donoghue, B.~R.~Holstein, L.~Planté and P.~Vanhove, ``Bending of Light in Quantum Gravity,''
 \href{\doibase 10.1103/PhysRevLett.114.061301}{  Phys.\ Rev.\ Lett.\  {\bf 114} (2015) no.6,  061301}.
 
 
 \bibitem{Younas:2015sva} 
  A.~Younas, S.~Hussain, M.~Jamil and S.~Bahamonde,
  ``Strong Gravitational Lensing by Kiselev Black Hole,''
  \href{\doibase 10.1103/PhysRevD.92.084042}{Phys.\ Rev.\ D {\bf 92}, 084042 (2015)}.
  
  
  \bibitem{Azreg-Ainou:2017obt} 
  M.~Azreg-Aïnou, S.~Bahamonde and M.~Jamil,
  ``Strong Gravitational Lensing by a Charged Kiselev Black Hole,''
  \href{\doibase 10.1140/epjc/s10052-017-4969-4}{Eur.\ Phys.\ J.\ C {\bf 77}, 414 (2017)}.

		%
		%
		%
		%
		%
		%
		%
		
		%
		%
		%
		%
		
		%
		%
		
		%
		%
		%
		%
		
		
		
	\end{thebibliography}
\end{document}